\documentclass{jpp}
\usepackage{graphicx,xcolor}

\usepackage{svg}
\usepackage[utf8]{inputenc}
\usepackage[T1]{fontenc}
\usepackage{amsmath}
\usepackage[colorlinks=true, allcolors=blue]{hyperref}

\shorttitle{Role of ion acoustic instability in magnetic reconnection}
\shortauthor{D. Li, Z. Liu and N. F. Loureiro}

\title{Role of ion acoustic instability in magnetic reconnection}

\author{Dion Li\aff{1}
  \corresp{\email{dionli@psfc.mit.edu}},
  Zhuo Liu\aff{1}
 \and Nuno F. Loureiro\aff{1}}

\affiliation{\aff{1}Plasma Science and Fusion Center, Massachusetts Institute of Technology, Cambridge, MA 02139, USA}

\begin{document}

\maketitle

\begin{abstract}
We report on a first-principles numerical study of magnetic reconnection in plasmas with different initial ion-to-electron temperature ratios.
In cases where this ratio is significantly below unity, we observe intense wave activity in the diffusion region, driven by the ion-acoustic instability.
Our analysis shows that the dominant macroscopic effect of this instability is to drive substantial ion heating. In contrast to earlier studies reporting significant anomalous resistivity, we find that anomalous contributions due to the ion-acoustic instability are minimal. These results shed light on the dynamical impact of this instability on reconnection processes, offering new insights into the fundamental physics governing collisionless reconnection.
\end{abstract}

\section{Introduction}\label{sec:introduction}
Magnetic reconnection is a phenomenon in which the topology of the magnetic field in a plasma is rearranged and magnetic energy is converted into kinetic or thermal energy \citep{Yamada2010,Ji2022}. Reconnection has been widely observed and has broad applications in space plasma events, such as solar flares \citep{Forbes1991} and coronal mass ejections \citep{Gosling1995}, the Earth’s and planetary magnetospheres \citep{Chen2008,Slavin2009,Le2017,Phan2018,Hesse2020}, and in a wide variety of astrophysical and laboratory settings \citep{Yamada1994,Yamada2010,Uzdensky2011,Hare2018}.

In the realm of collisionless reconnection, a persistent challenge has been to pinpoint the kinetic processes that facilitate rapid topological changes and energy conversion. Specifically, kinetic instabilities have been proposed to play an important role. These instabilities, through intense wave-particle interactions, may generate anomalous resistivity \citep{Sagdeev1967,1983bpp..conf..389G,Labelle1988,Treumann2014,Liu2024}, and could determine the reconnection rate \citep{Ji1998,Kulsrud1998,Kulsrud2014,Uzdensky2003,Treumann2014}. Observational evidence \citep{Deng2001,Farrell2002,Matsumoto2003,Eastwood2009,Laitinen2010,Khotyaintsev2019,Cozzani2023} and numerical simulations \citep{Drake2003,JaraAlmonte2014,Fujimoto2014,Le2018,Dokgo2019,Le2019,Ng2020a,Wang2021,Zhang2023,Yi2023} confirm the presence of turbulent phenomena and wave emissions in the vicinity of diffusion regions at reconnection sites, indicating that these instabilities are active players in  reconnection dynamics. Moreover, they might affect the onset of reconnection itself \citep{Ricci2004,Alt2019,Winarto2022}.

Among the various potential kinetic modes, the ion-acoustic instability (IAI), driven by the relative drift between ions and electrons (or the electric current), has been proposed as a key factor in the dissipation of magnetic energy in collisionless plasmas \citep{Coppi1971,Smith1972,Coroniti1977,Sagdeev1979} and appears to be a compelling explanation for numerous observations of ion-acoustic waves (IAWs) across different plasma settings. As far back as the 1970s, the Helios I and II missions have documented the presence of IAWs within heliocentric distances ranging from 0.3 to 1 AU \citep{Gurnett1977,Gurnett1978}, setting the stage for renewed interest in the role of these waves in the solar wind. More recently, advanced instrumentation from NASA's Parker Solar Probe (PSP) \citep{Fox2015} and the European Space Agency's Solar Orbiter \citep{Mller2012} has further highlighted the prominence of IAWs in the near-Sun environment. For example, the Solar Orbiter’s Time Domain Sampler receiver reveals that IAWs are a dominant wave mode close to the Sun \citep{Graham2021,Pa2021}, underscoring their importance for solar and heliospheric physics.

Observations of these waveforms may shed light on a longstanding question in solar physics: why proton temperatures in the solar wind decline much more slowly than predicted by simple adiabatic expansion, especially at distances beyond $20$ AU \citep{Bridge1977,Richardson2003,Mozer2023}. This discrepancy implies the presence of additional heating mechanisms, possibly due to the turbulent nature of the solar wind \citep{Tu1995,Bruno2005}. In particular, turbulence can drive localized magnetic reconnection events \citep{Retin2007,Servidio2009,Servidio2011,Osman2014,Loureiro2017} in the solar wind, with associated large currents. The existence of solar wind patches where ions are noticeably colder than electrons \citep{Hellinger2013,tverk2015,Chen2016,Verscharen2019,Pa2021} suggests that IAI can potentially be triggered in such events. Recent studies \citep{Liu2024} and our own findings in this paper suggest that IAI can efficiently heat ions, providing a plausible explanation (among others) for the persistent, but not fully understood, ion heating in the solar wind over vast distances.

In this investigation, we focus specifically on the role of IAI in a simple, two-dimensional reconnecting configuration to determine whether it can provide a viable pathway for ion heating and wave generation within collisionless reconnection. Through a series of high-fidelity particle-in-cell numerical simulations, we find that this is indeed the case when the initial ion temperature is significantly lower than that of the electrons.

This paper is organized as follows. Section~\ref{sec:theoretical_framework} outlines the theoretical framework of IAI, highlighting its relevance to magnetic reconnection. Section~\ref{sec:simulation_setup} describes the particle-in-cell simulation setup and parameters. Section~\ref{sec:simulation_results} presents the results, including the reconnection rate (\ref{sec:rec_rate}), conditions for IAI emergence (\ref{sec:onset}), evidence of IAW generation (\ref{sec:identification_of_iaws}), and the role of IAI in reconnection dynamics (\ref{sec:effects_in_magnetic_reconnection}). Finally, Section~\ref{sec:conclusions} discusses the implications of these findings on ion heating in the solar wind and suggests directions for future research.

\section{Theoretical framework}\label{sec:theoretical_framework}
In this section, we revisit the IAI derivation by examining the wave mode solution parallel to the electron-ion drift. For non-relativistic systems, the Vlasov-Poisson equations are expressed as 
\begin{align}
\frac{\partial f_{\alpha}}{\partial t} + \mathbf{v} \cdot \nabla f_{\alpha} & + \bigg( - \frac{q_{\alpha}}{m_{\alpha}} \nabla \varphi \bigg) \cdot \frac{\partial f_{\alpha}}{\partial \mathbf{v}} = 0, \label{subeq:VP1} \\
\nabla^2 \varphi &= -4\pi \sum_{\alpha} q_{\alpha} \int d^3 v~f_{\alpha}, \label{subeq:VP2}
\end{align}
with $\alpha=i,e$ the particle species index. The combined and linearized equations for the $1+1$-dimensional $(x,v_x)$ case can be expressed in the form of a dielectric function
\begin{align}\label{eq:linVlasov}
\epsilon (p, k) = 1 - \sum_{\alpha} \frac{\omega_{p\alpha}^2}{k^2} \frac{1}{n_{\alpha}} \int dv_x~\frac{F^{\prime}_{\alpha} (v_x)}{v_x - ip/k},
\end{align}
where $\omega_{p\alpha}$ is the plasma frequency, $p \equiv -i\omega + \gamma$, and $F^{\prime}_{\alpha} (v_x)$ is the derivative of the 1D velocity distribution function with respect to $v_x$. In the electrostatic scenario, the $l.h.s.$ of Eq.~\ref{eq:linVlasov} is identically zero. 
By assuming both ion and electron species follow Maxwellian distributions, with electrons drifting at a speed $U_d$ relative to the ions, Eq.~\ref{eq:linVlasov} yields
\begin{align}\label{eq:solVlasov}
1 + \frac{1 + Z_p (\frac{\omega - kU_d}{\sqrt{2} k v_{\text{th},e}})}{k^2 \lambda_{De}^2} + \frac{1 + Z_p (\frac{\omega}{\sqrt{2} k v_{\text{th},i}})}{k^2 \lambda_{Di}^2} & \nonumber \\
+ i \bigg[ \frac{Z_{\Delta} (\frac{\omega - kU_d}{\sqrt{2} k v_{\text{th},e}})}{k^2 \lambda_{De}^2} + \frac{Z_{\Delta} (\frac{\omega}{\sqrt{2} k v_{\text{th},i}})}{k^2 \lambda_{Di}^2} & \bigg] = 0,
\end{align}
where $v_{\text{th},\alpha} = \sqrt{T_{\alpha}/m_{\alpha}}$ represents the thermal velocity, $\lambda_{D\alpha}$ the Debye length, and we define
\begin{align}
Z_p (\zeta) & \equiv - \text{erfi} (\zeta) Z_{\Delta} (\zeta), \label{subeq:Vdef1} \\
Z_{\Delta} (\zeta) & \equiv \sqrt{\pi} \zeta e^{-\zeta^2}, \label{subeq:Vdef2}
\end{align}
and $\text{erfi} (\zeta)$ the imaginary error function. Equation~\ref{eq:solVlasov} can be solved numerically to obtain the growth rates and spectrum of IAWs. Notably, IAI is only triggered under specific conditions. For $T_{i}/T_{e} \ll 1$, a necessary criterion is that the electron-ion drift speed must be approximately or larger than the ion-sound speed $c_s = \sqrt{(\gamma_e T_e + \gamma_i T_i)/m_i} \approx \sqrt{(T_e + 3 T_i)/m_i}$\footnote{For $T_i/T_e < 1/20$, the critical drift velocity for triggering IAI is actually $U_{d,\text{IAI}} \simeq 4\sqrt{2} v_{\text{th},i}$ \citep{Fried1961}. Only at higher temperature ratios (e.g., $T_i/T_e > 1/10$) does this threshold become comparable to $c_s$ (or larger than $c_s$ as the temperature ratio approaches or exceeds unity). In our coldest‐ion case, however, $T_{i0}/T_{e0} = 1/50$ and $4\sqrt{2} v_{\text{th},i}/c_s \simeq 4\sqrt{2} \sqrt{T_i/T_e} = 0.8 \sim 1$, which justifies the approximation that the threshold is $\sim c_s$.}, where $\gamma_{\alpha}$ is the adiabatic index of species $\alpha$ and we have assumed that electrons are isothermal ($\gamma_e = 1$) and ions are adiabatic ($\gamma_i = 3$) \citep{Biskamp2000}.

In the context of collisionless magnetic reconnection, one may expect IAI to be driven in both the out-of-plane and in-plane directions. The out-of-plane direction is influenced by the reconnection current and the associated reconnection electric field, while the in-plane direction is driven by the current and electric field resulting from decoupled ion and electron flows in the outflow direction. The former mechanism was briefly addressed in \citet{Liu2024}. The condition for the onset of out-of-plane IAI can be approximated as
\begin{align}
\frac{J_z/en}{c_s} \sim \frac{cB_0}{4\pi \delta en} \sqrt{\frac{m_i}{T_e}} = \sqrt{\frac{2}{\beta_e}} \frac{d_i}{\delta} > 1, \label{eq:out-of-plane_condition}
\end{align}
where $J_z$ is the out-of-plane current density, $B_0$ is the upstream (reconnecting) magnetic field, $\delta$ is the current sheet thickness, $d_{i} = c/\omega_{pi} = c (m_i/4\pi n e^2)^{1/2}$ is the ion inertial length or ion skin-depth, and $\beta_e$ is the electron plasma beta based on the reconnecting field. This condition is easy to satisfy since $\delta$ is expected to be on the order of the electron skin-depth.

\begin{figure}
\begin{center}
\includegraphics[width=0.7\textwidth]{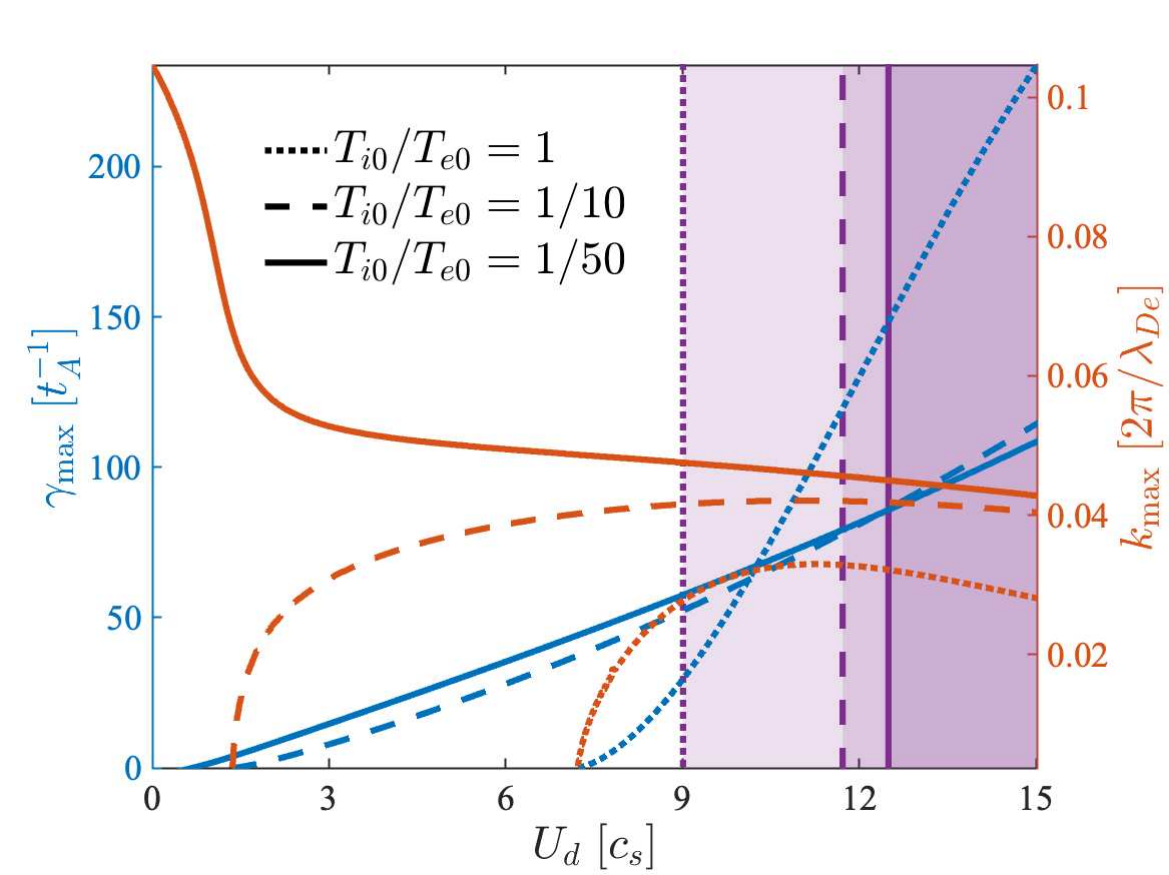}
\caption{\label{fig:pk} Maximum growth rates ($\gamma_{\text{max}}$, in blue) and the corresponding wavenumbers ($k_{\text{max}}$, in orange) as functions of drift velocity ($U_d$) for $T_{i0}/T_{e0} = 1$ (dotted lines), $T_{i0}/T_{e0} = 1/10$ (dashed lines), and $T_{i0}/T_{e0} = 1/50$ (solid lines). These values are obtained from solving Eq.~\ref{eq:solVlasov}, using the thermal velocities initialized in the simulations and $n_i = n_e = n_b = 0.2 n_0$. Regions not accessible within the considered reconnection configuration along the outflow symmetry line are shaded in purple, with the purple lines representing the maximum expected electron-ion drift velocity ($U_{d,\text{in-plane}} = v_{A,e} - v_{A,i}$) at the ends of the diffusion region for each ion-electron temperature ratio. The increase in $k_{\text{max}}$ at $\gamma_{\text{max}} = 0$ as $T_{i0}/T_{e0}$ decreases beyond $1/10$ is explained in Appendix~\ref{appA}. }
\end{center}
\end{figure}

Likewise, it is inevitable that the in-plane IAI will be triggered if $T_i/T_e \ll 1$ because the in-plane electron-ion drift should reach $U_{d,\text{in-plane}} \approx v_{A,e} - v_{A,i}$, giving
\begin{align}
\frac{U_{d,\text{in-plane}}}{c_s} \sim \frac{v_{A,e} - v_{A,i}}{c_s} \approx \frac{v_{A,e}}{c_s} = \frac{B_0}{\sqrt{4\pi n_e m_e}} \sqrt{\frac{m_i}{T_e}} = \sqrt{\frac{m_i}{m_e}} \sqrt{\frac{2}{\beta_e}} \gg 1, \label{eq:in-plane_condition}
\end{align}
where $v_{A,\alpha} = B_0/(4\pi n_{\alpha} m_{\alpha})^{1/2}$ is the Alfv\'en speed of species $\alpha$. This result is illustrated in Fig.~\ref{fig:pk}, which presents numerical solutions of the growth rate for the fastest growing wave mode and the wavenumber at which this maximum growth rate occurs, calculated by solving Eq.~\ref{eq:solVlasov} using the initial thermal velocities from the simulations that follow and assuming uniform number densities $n_i = n_e = n_b = 0.2 n_0$ (the approximate ion and electron density near the x-point around the time of peak reconnection rate in all simulations). The figure shows that there exists a range of drift speeds below the theoretical maximum of $U_d = v_{A,e} - v_{A,i}$ (identified by the purple vertical lines in Fig.~\ref{fig:pk}\footnote{Notice the reason that these lines do not overlap in Fig.~\ref{fig:pk} is imposed by the requirements of the equilibrium setup (see Section~\ref{sec:simulation_setup}), resulting in different $T_{e0} + 3 T_{i0}$ (and hence, $c_s$) for different temperature ratios.}) for which the growth rate of the fastest growing mode is positive for all ion-electron temperature ratios considered. In the numerical simulations that follow, we find that outflow electron-ion drift speed $|U_{d,\text{outflow}}| = |u_{x,e} - u_{xi}|$ reaches $\approx v_{A,e}/2$ at the outflow ends of the diffusion region, a result that is roughly independent of temperature ratio (see Figs.~\ref{fig:ps} and~\ref{fig:Ud_evol}), which, alongside Fig.~\ref{fig:pk}, suggests that IAI can be strongly triggered for $T_{i0}/T_{e0} = 1/10$ and $1/50$, but not for $T_{i0}/T_{e0} = 1$. 

Due to computational constraints, this study considers only the two-dimensional case, with the out-of-plane and invariant direction aligned with $\hat{\mathbf{z}}$, and focuses exclusively on the in-plane component of IAI.

\section{Simulation setup}\label{sec:simulation_setup}
The simulations reported in this work are conducted using the particle-in-cell (PIC) code \textsc{osiris} \citep{Fonseca2002,Hemker2015}. The initial conditions are set to a Harris sheet equilibrium \citep{Harris1962}, with the initial magnetic vector potential expressed as
\begin{eqnarray}
\textbf{A} = \hat{\textbf{z}} B_0 \lambda_{\text{B}} \ln \bigg[ \cosh \bigg( \frac{y}{\lambda_{\text{B}}} \bigg) \bigg].
\end{eqnarray}
Here, $\lambda_{\text{B}}$ is the characteristic length scale of the magnetic field gradient (half-width of the current sheet). The initial density distribution is defined as 
\begin{eqnarray}
n_e = n_i = n_0 \text{sech}^2 \bigg( \frac{y}{\lambda_{\text{B}}} \bigg) + n_b,
\end{eqnarray}
with $n_b = 0.2n_0$ as the background number density.

The global Alfv\'en time is defined as $t_A = L_x/v_{A,i0}$, where $L_x$ and $L_y$ are the lengths of the simulation box in the $x$ (outflow) and $y$ (inflow) directions. The ion and electron out-of-plane current densities are specified by $J_{zi0}/J_{ze0} = T_{i0}/T_{e0}$ to establish Vlasov equilibrium \citep{Schindler2006}, with $T_{i0}$ and $T_{e0}$ uniform, and $\mathbf{J}_0 = c(\nabla \times \mathbf{B}_0/4\pi)$. The force-balance condition is satisfied by setting $n_0 (T_{i0} + T_{e0}) = B_0^2/8\pi$. The relationship between the electron plasma and cyclotron frequencies is given by $\omega_{pe} = 2\Omega_{ce}$ where $\omega_{pe} = (4\pi n_0 e^2/m_e)^{1/2}$ and $\Omega_{ce} = |e|B_0/(m_e c)$.

The initial perturbation is similar to that described in \citep{Daughton2009_1}, with $\delta \mathbf{B} = \hat{\mathbf{z}} \times \nabla \psi$ and $\psi = - \delta \psi \cos (2\pi x/L_x) \cos (\pi y/L_y)$, where the perturbation amplitude is set to $\delta \psi/B_0 d_{i0} = 0.1$. We utilize cubic particle interpolations and resolve the electron Debye length ($\lambda_{De}$) to prevent numerical heating. As a result, it was essential to adjust the resolution and timestep parameters based on the ion-electron temperature ratio.

Key simulation parameters used in these runs are listed in Tables~\ref{tab:table1} and~\ref{tab:table2}.

\begin{table}
\begin{center}
\begin{tabular}{cccc}
& Parameter & Value & \\
\hline
& $8 \pi n_0 (T_{i0} + T_{e0})/B_0^2$ & $1$ & \\
& $\omega_{pe}/\Omega_{ce}$ & $2$ & \\
& $m_i/m_e$ & $100$ & \\
& $L_x/d_{i0}$ & $32$ & \\
& $\lambda_{\text{B}}/d_{i0}$ & $1$ & \\
& $n_b/n_0$ & $0.2$ & \\
& $t_{\text{max}}/t_{A}$ & $2.25$ & \\
& $\delta \psi/B_0 d_{i0}$ & $0.1$ & \\
\hline
& Particles-per-cell & $256~(16 \times 16)$ & \\
\end{tabular}
\caption{Summary of simulation parameters.}
\label{tab:table1}
\end{center}
\end{table}

\begin{table}
\begin{center}
\begin{tabular}{cccc}
$T_{i0}/T_{e0}$ & $1/50$ & $1/10$ & $1$ \\
\hline
$L_y/d_{i0}$ & $24$ & $24$ & $16$ \\
$\Delta x/\lambda_{De} = \Delta y/\lambda_{De}$ & $0.321$ & $\sim 1/3$ & $\sim 1/3$ \\
$\lambda_{De}/d_{e0} = u_{\text{th}, e}/c$ & $0.350$ & $0.337$ & $0.250$  \\
\hline
Resolution & $2848 \times 2136$ & $2848 \times 2136$ & $3840 \times 1920$ \\
Timestep ($\omega_{pe} \Delta t$) & $0.0786$ & $0.0786$ & $0.0583$ \\
\end{tabular}
\caption{Simulation-specific parameters.}
\label{tab:table2}
\end{center}
\end{table}

\section{Simulation results and wave activity in the diffusion region}\label{sec:simulation_results}
In this section we describe and analyze in detail the results of our numerical simulations. We begin with an overview of the reconnection dynamics observed across the simulations, emphasizing the relatively minor impact of the temperature ratio on the reconnection rate. Despite this, significant wave activity is detected in the diffusion region, leading to pronounced ion heating. We argue in the sections below that this wave activity is driven by IAI, and characterize its onset and impact on reconnection dynamics.

\subsection{Reconnection rate}\label{sec:rec_rate}
\begin{figure}
\begin{center}
\includegraphics[width=0.7\textwidth]{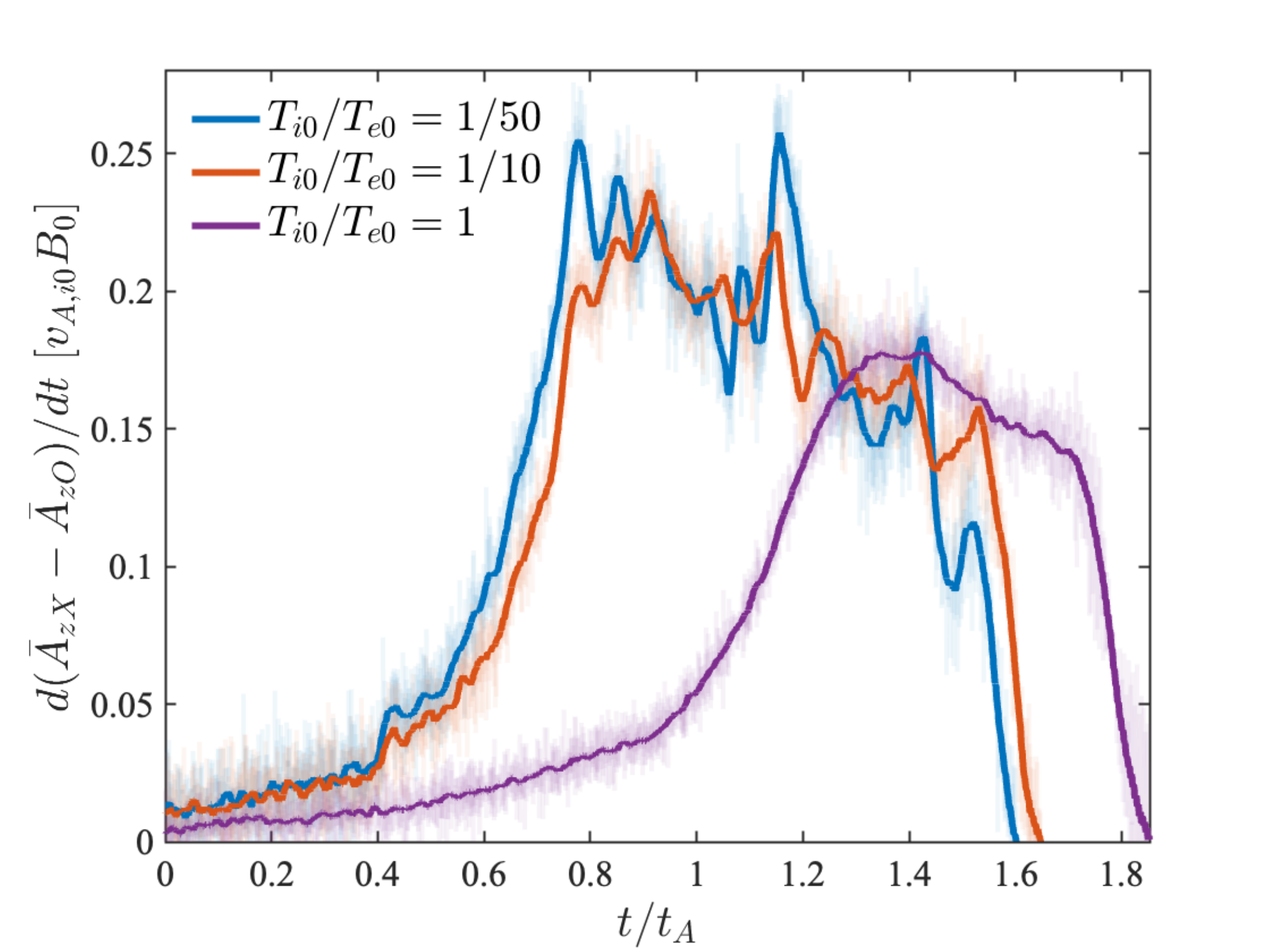}
\caption{\label{fig:ER_max} Time evolution of the temporally averaged reconnection rate for $T_{i0}/T_{e0} = 1/50$ (blue), $T_{i0}/T_{e0} = 1/10$ (orange), and $T_{i0}/T_{e0} = 1$ (purple). $A_{zX}$ and $A_{zO}$ are the out-of-plane magnetic vector potentials at the X and O points, respectively, and the semi-transparent curves are more exact values of the solid curves. }
\end{center}
\end{figure}

Plotted in Fig.~\ref{fig:ER_max} is the normalized reconnection rate (computed as the difference in the magnetic vector potential's rate of change at the X and O points, and temporally averaged over a moving window of $\approx 0.03~t_A$) for each value of the ion-electron temperature ratio that we have numerically simulated. 
As shown, we find that the onset of fast reconnection occurs earlier in time for colder ions; however, the peak reconnection rate is weakly dependent on the temperature ratio—varying by less than a factor of 2 across values of $T_{i0}/T_{e0}$ ranging from $1/50$ to $1$—and is consistent with the established values in the literature~\citep[e.g.,][]{Birn2001,Cassak2017}. 
Also evident from the plot is the highly oscillatory nature of the curves corresponding to low initial temperature ratios during the fast reconnection stage, in contrast with the $T_{i0}/T_{e0}=1$ case. 
This behavior hints at turbulent dynamics in the current sheet during those times which, as we will demonstrate below, is due to the destabilization of ion-acoustic modes.

\subsection{Onset of Ion Acoustic Instability (IAI)}\label{sec:onset}
\begin{figure*}
\includegraphics[width=1.03\textwidth,scale=.5]{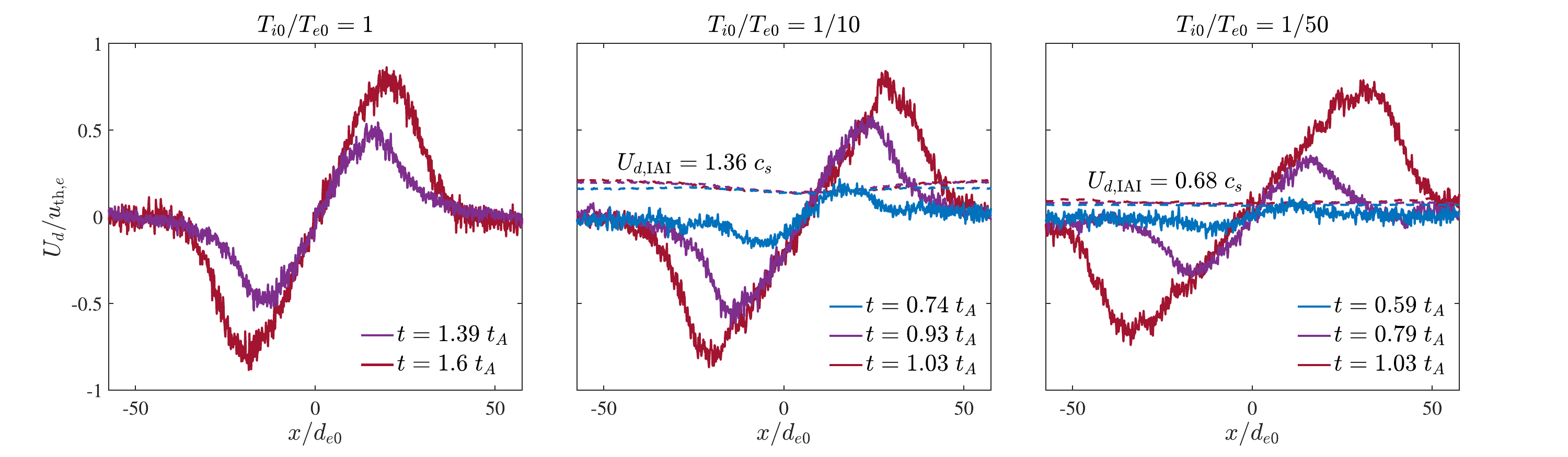}
\caption{\label{fig:Ud_evol} Evolution of the outflow electron-ion drift ($U_d = u_{x,e} - u_{x,i}$), normalized to the initial electron thermal velocity $u_{\text{th},e}$, along the outflow symmetry line ($y = 0$) for $T_{i0}/T_{e0} = 1$ (left), $T_{i0}/T_{e0} = 1/10$ (middle), and $T_{i0}/T_{e0} = 1/50$ (right). The blue curves denote the times when the drifts begin to exceed the corresponding IAI thresholds (which, for $T_{i0}/T_{e0} = 1$ never occurs and thus no blue curve is shown in the left plot), the purple curves refer to the approximate times of peak reconnection rates, and the red curves represent later times when the drifts reach maximum values. Dashed lines show the corresponding IAI threshold drift speeds, calculated with local ion-acoustic speeds $c_s = \sqrt{(T_{e,xx} + 3 T_{i,xx})/m_i}$. }
\end{figure*}

We find that the outflow electron-ion drift speed $|U_{d,\text{outflow}}|$ exceeds the IAI threshold drift in the diffusion region and near the separatrix for the cases $T_{i0}/T_{e0}=1/50$ and $1/10$. 
This is shown for the diffusion region in Fig.~\ref{fig:Ud_evol}, which illustrates the evolution of the outflow electron-ion drift along the outflow symmetry line ($y = 0$) near the x-point for the different temperature ratios considered. This drift occurs within the ion diffusion region because ions decouple from the magnetic field lines on electron scales, allowing for both ion and electron outflow speeds to approach their respective Alfv\'en speeds \citep{Shay1998,Liu2022}.

\begin{figure*}
\includegraphics[width=1\textwidth,scale=.5]{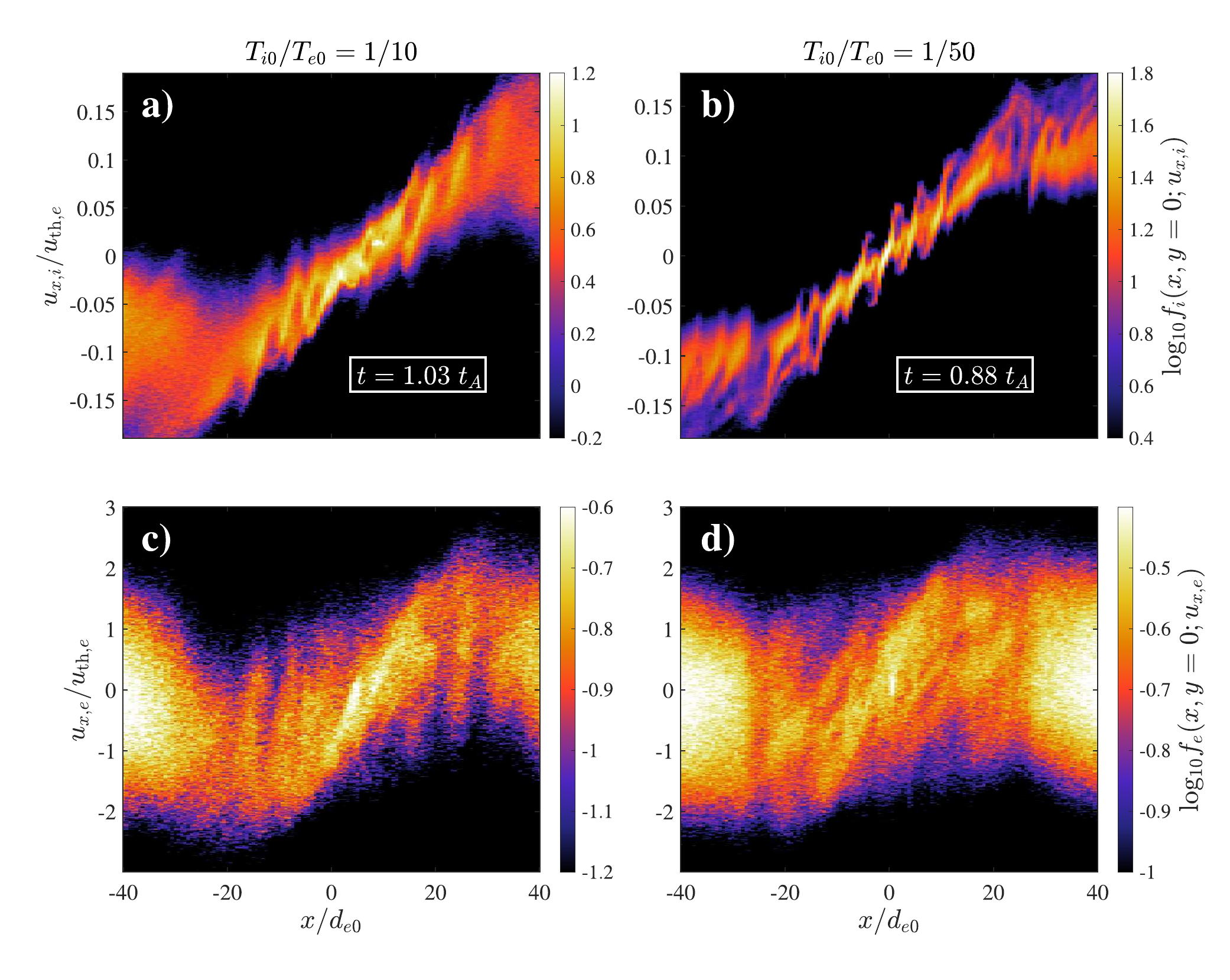}
\caption{\label{fig:ps} Ion (top row) and electron (bottow row) phase-space distributions in $u_x$ and $x$ along the outflow symmetry line ($y = 0$) near the x-point for $T_{i0}/T_{e0} = 1/10$ (left column) and $T_{i0}/T_{e0} = 1/50$ (right column) at the approximate times when IAI is strongly triggered. }
\end{figure*}

\begin{figure*}
\includegraphics[width=1.0\textwidth,scale=.5]{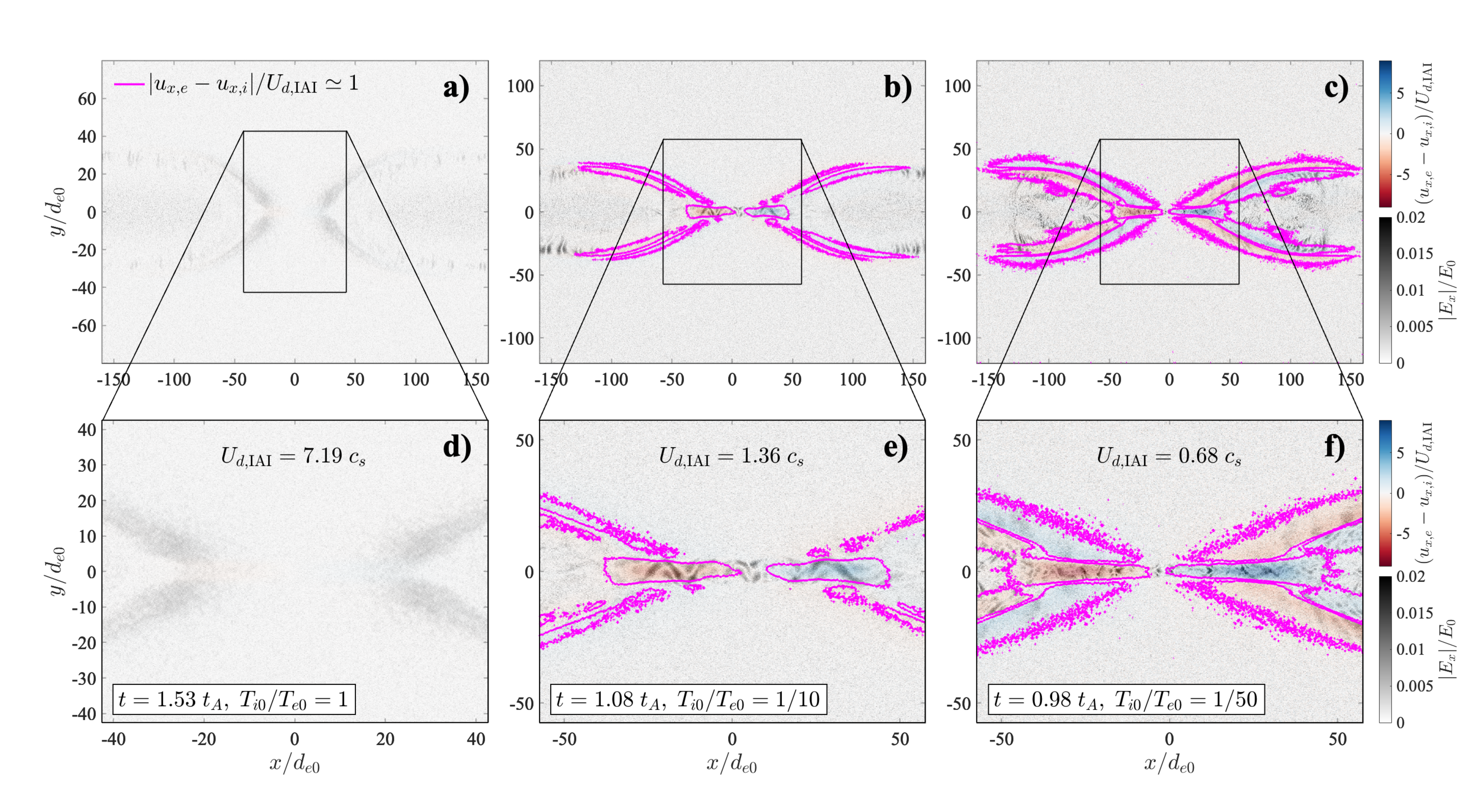}
\caption{\label{fig:comp} Colormap displaying the $x$-component of the electron-ion drift velocity normalized to the IAI threshold drifts, calculated with local ion-acoustic speeds $c_s = \sqrt{(T_{e,xx} + 3 T_{i,xx})/m_i}$ (red-blue), overlaid with contours indicating the regions where $|u_{x,e} - u_{x,i}|/U_{d,\text{IAI}} \simeq 1$ (magenta) and the magnitude of the outflow electric field ($E_x$, normalized to the reference value $E_0 = m_e c \omega_{pe}/e$ and in gray-scale) shortly after the time of maximum reconnection rate (see Fig.~\ref{fig:ER_max}). This is shown for the following cases: $T_{i0}/T_{e0} = 1$ (left-most column), $T_{i0}/T_{e0} = 1/10$ (middle column), and $T_{i0}/T_{e0} = 1/50$ (right-most column). 
The bottom row figures zoom in on the diffusion region, as identified by the boxes in the top row. }
\end{figure*}

The separation of electron and ion outflow velocities is illustrated in the phase-space distributions presented in Fig.~\ref{fig:ps} (see also Fig.~\ref{fig:Ud_evol} and the contours in Fig.~\ref{fig:comp}), which shows ion and electron phase-space densities as functions of the particles' proper velocity and position in the outflow direction ($u_x$ and $x$, respectively). The upper row of Fig.~\ref{fig:ps} also reveals nonlinear ion phase-space structures, which coincide with the emergence of waveforms in the diffusion region that, in the next subsection, we associate with IAWs. 

Notably, Fig.~\ref{fig:ps} demonstrates that the maximum ion proper velocities (which are approximately ion bulk velocities due to the relatively low ion temperatures), attained at the end-points of the current sheet, closely match the Alfv\'en speed at those locations ($v_{A,i} \simeq 0.15 u_{\text{th,e}}$), aligning with the established understanding of reconnection physics that the bulk plasma is accelerated to $v_{A,i}$ in the outflow direction. The maximum electron outflow velocity, which is $\approx 0.8 u_{\text{th,e}} \approx v_{A,e}/2$ (see Fig.~\ref{fig:Ud_evol}), being much larger than that of the ions implies that there's a large in-plane current. \footnote{We can analyze the development of the electron-ion drift in the diffusion region by calculating the expected proper velocity for each species at each $x$-position near the x-point using the formula $\langle u_{x,\alpha} \rangle = \int du_{x,\alpha}~u_{x,\alpha} f(x, y = 0, u_{x,\alpha})/\int du_{x,\alpha}~f(x, y = 0, u_{x,\alpha})$. Here, $u$ is used to calculate the drift instead of $v$ because, formally, the vertical dimension of the phase-space distributions shown in Fig.~\ref{fig:ps} represents the proper velocity vectors $\mathbf{u} = \gamma \mathbf{v}$. We observe that the electron proper velocity in the $x$-direction near the outflow edges can reach up to $u_{x,e} \simeq 0.7c$, corresponding to a Lorentz factor of $\gamma \simeq 1.22$. This value indicates that the motion is only mildly relativistic, thereby justifying the use of the nonrelativistic Vlasov-Poisson equations Eqs.~\ref{subeq:VP1} and~\ref{subeq:VP2} for interpreting the results.}

Figure~\ref{fig:Ud_evol} shows that that the electron-ion drifts begin to exceed the corresponding IAI thresholds (see Fig.~\ref{fig:pk}, which gives $U_{d,\text{IAI}}/c_s = U_d (\gamma_{\text{max}} = 0)/c_s \approx 7.19$, $1.36$, and $0.68$ for $T_{i0}/T_{e0} = 1$, $1/10$, and $1/50$) at $t \simeq 0.74~t_A$ and $t \simeq 0.59~t_A$ for $T_{i0}/T_{e0} = 1/10$ and $1/50$, respectively. These approximate times correlate well with the emergence of waveforms in the diffusion region (see Fig.~\ref{fig:ss_y0} for the $T_{i0}/T_{e0} = 1/50$ case).

We further demonstrate the decoupling between electron and ion drifts in Fig.~\ref{fig:comp}, which displays colormaps of the outflow electron-ion drift velocity along with contours highlighting where this drift exceeds the associated IAI threshold drift. Additionally, it depicts the magnitude of the outflow electric field ($E_x$), whose wave-like structures we examine in detail in the next section.

\subsection{Identification of IAWs}\label{sec:identification_of_iaws}
A salient feature in the contour plots of Fig.~\ref{fig:comp} is the appearance of wave-like structures in the diffusion region, particularly for the cases where $T_{i0}/T_{e0} = 1/10$ and $1/50$. These structures suggest the presence of underlying kinetic instabilities, likely associated with IAWs. To better understand the origin of this wave activity, we now focus on a detailed analysis of the diffusion region.

To isolate the waveforms specifically associated with the instability, we concentrate on the outflow electric field, $E_x$. This approach helps to avoid interference from background electromagnetic field structures, as the baseline outflow electric field is much weaker than the observed fluctuations. As a result, any detected $E_x$ perturbations can be directly attributed to wave activity. Figure~\ref{fig:comp} highlights the regions where large-amplitude waveforms of the outflow electric field are concentrated, indicating that these waves are generated and confined in regions where the outflow electron-ion drift velocity significantly exceeds the IAI threshold drift. These observations, combined with the absence of significant wave activity in the $T_{i0}/T_{e0} = 1$ case, strongly suggest that the waveforms are generated by IAI, triggered by the large outflow current.

\begin{figure*}
\includegraphics[width=1\textwidth,scale=.5]{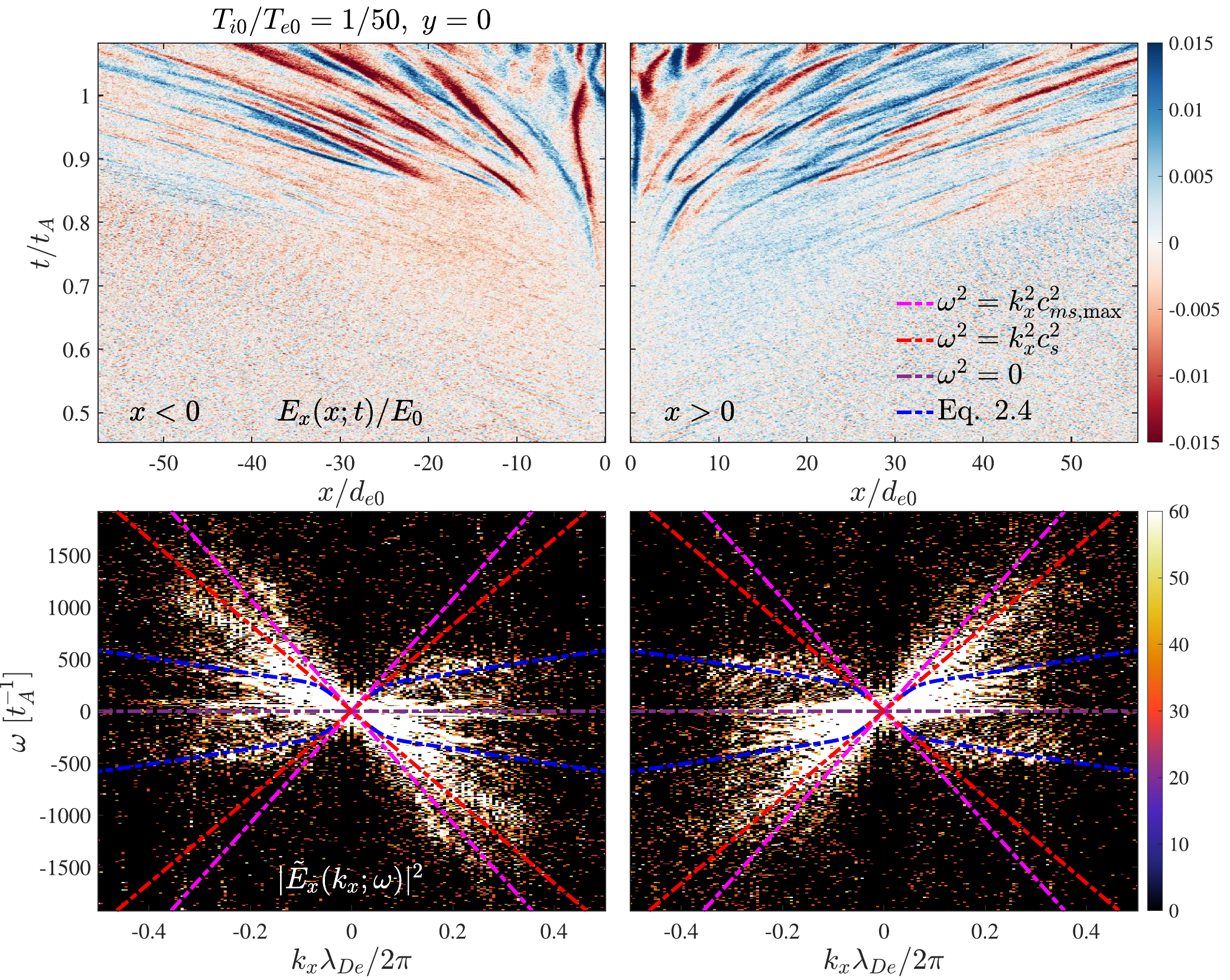}
\caption{\label{fig:ss_y0} Outflow electric field component ($E_x$) near the x-point along the outflow symmetry line ($y = 0$) as a function of $x$ and $t$ for $T_{i0}/T_{e0} = 1/50$ (top row) and the squared two-dimensional spectrum $|\tilde{E} (k_x; \omega)|^2$ (bottom row) of the $x < 0$ (left column) and $x > 0$ (right column) halves of the data. The spectrum is overlaid with the MHD compressional mode solutions (Eqs.~\ref{subeq:MHDsol2} in purple,~\ref{subeq:MHDsol1a} in magenta, and~\ref{subeq:MHDsol1b} in red, using the initial ion and electron temperatures to calculate $c_s$) for all $k_x$, though the MHD compressional mode (Eq.~\ref{eq:MHDcomp}) is strictly valid for $|k_x| \ll d_{i}^{-1}$. Additionally, the parallel wave mode solution to the linearized Vlasov-Poisson equations (Eq.~\ref{eq:solVlasov}, in blue) is shown for an electron-ion drift speed of $U_d = v_{A,e} - v_{A,i}$, corresponding to the maximum approximate drift near the outflow edges. It should be noted that blue curves from Eq.~\ref{eq:solVlasov} remain largely unchanged for different $U_d$ as computations over a range of $U_d$ confirm that the real frequency depends only weakly on this parameter. }
\end{figure*}

The evolution of these waveforms for $T_{i0}/T_{e0} = 1/50$ is depicted in the top row of Fig.~\ref{fig:ss_y0}, showing the outflow electric field near the x-point along the ouflow symmetry line ($y = 0$) as a function of $x$ and $t$. By performing two-dimensional Fourier transforms, we obtain spectra of the outflow electric field near the x-point, shown in the bottom row of Fig.~\ref{fig:ss_y0}. These spectra are compared with the solutions to the linearized Vlasov-Poisson equations (Eq.~\ref{eq:solVlasov}) and the MHD compressional mode, given by~\citep[e.g.][]{Bellan2006}
\begin{align}\label{eq:MHDcomp}
(\omega^2 - k^2 c_s^2) (k^2 v_{A,i}^2 - \omega^2) + k^2 v_{A,i}^2 k_{\perp}^2 c_s^2 = 0,
\end{align}
where $k^2 \equiv k_{\parallel}^2 + k_{\perp}^2$. Along $y = 0$ the magnetic field is fully aligned with $y$, allowing us to write $k_{\parallel} = k_y$. For $k_y = k_z = 0$, the solutions to Eq.~\ref{eq:MHDcomp} are
\begin{align}
\omega^2 & = k_x^2 c_{ms}^2, \label{subeq:MHDsol1} \\
\omega^2 & = 0, \label{subeq:MHDsol2}
\end{align}
where we have introduced the magnetosonic speed $c_{ms} = (v_{A,i}^2 + c_s^2)^{1/2}$. Equation~\ref{subeq:MHDsol2} represents a ``non-propagating'' or ``static'' solution to the MHD compressional mode. 

Since $|B_y|$ increases from zero at the x-point to some maximum value at the ends of the current sheet (which we find in our simulations to be approximately $60\%$ of the far asymptotic field $B_0$ for the times considered in Fig.~\ref{fig:ss_y0}, thus giving $v_{A,i,\text{max}} \approx 0.6 v_{A,i0}$), we expect there to exist a range of waveforms between the curves 
\begin{align}
\omega^2 & = k_x^2 c_{ms,\text{max}}^2, \label{subeq:MHDsol1a} \\
\omega^2 & = k_x^2 c_s^2, \label{subeq:MHDsol1b}
\end{align}
where $c_{ms,\text{max}}^2 = v_{A,i,\text{max}}^2 + c_s^2$. We observe that the resultant spectra for $T_{i0}/T_{e0} = 1/50$ consist of structures that align closely with Eqs.~\ref{subeq:MHDsol2},~\ref{subeq:MHDsol1a}, and~\ref{subeq:MHDsol1b} and with the linearized Vlasov-Poisson equations (Eq.~\ref{eq:solVlasov}). This indicates that the observed outflow electric field waveforms in the top row of Fig.~\ref{fig:ss_y0} are likely due to the interference resulting from the superposition of IAWs and magnetosonic waves. Although the MHD compressional mode is only strictly valid for $|k_x| \ll d_{i}^{-1}$, Eqs.~\ref{subeq:MHDsol2},~\ref{subeq:MHDsol1a}, and~\ref{subeq:MHDsol1b} were plotted in Fig.~\ref{fig:ss_y0} for all $k_x$. Indeed, the spectra in Fig.~\ref{fig:ss_y0} begins to deviate from these solutions at high wavenumbers.

To demonstrate the bidirectional nature of IAWs, we analyzed the spectra from regions either to the left ($x < 0$) or to the right ($x > 0$) of the x-point (see Fig.~\ref{fig:ss_y0}, left and right bottom plots, respectively). In this analysis, we observe that large-amplitude waveforms propagating with the bulk ion plasma, which align with the MHD compressional mode (Eq.~\ref{eq:MHDcomp}) components of the spectrum, flow unidirectionally away from the x-point. In contrast, the lower phase velocity modes at high $k$, associated with the solutions to Eq.~\ref{eq:solVlasov} and indicative of IAWs, display both positive and negative slopes in frequency space, confirming their bidirectional nature. This evidence supports our assertion that IAWs generated near the x-point at locations where the in-plane drift is largest, corresponding to the fastest growing modes of IAI, propagate both toward and away from the x-point.

\begin{figure*}
\includegraphics[width=1\textwidth,scale=.5]{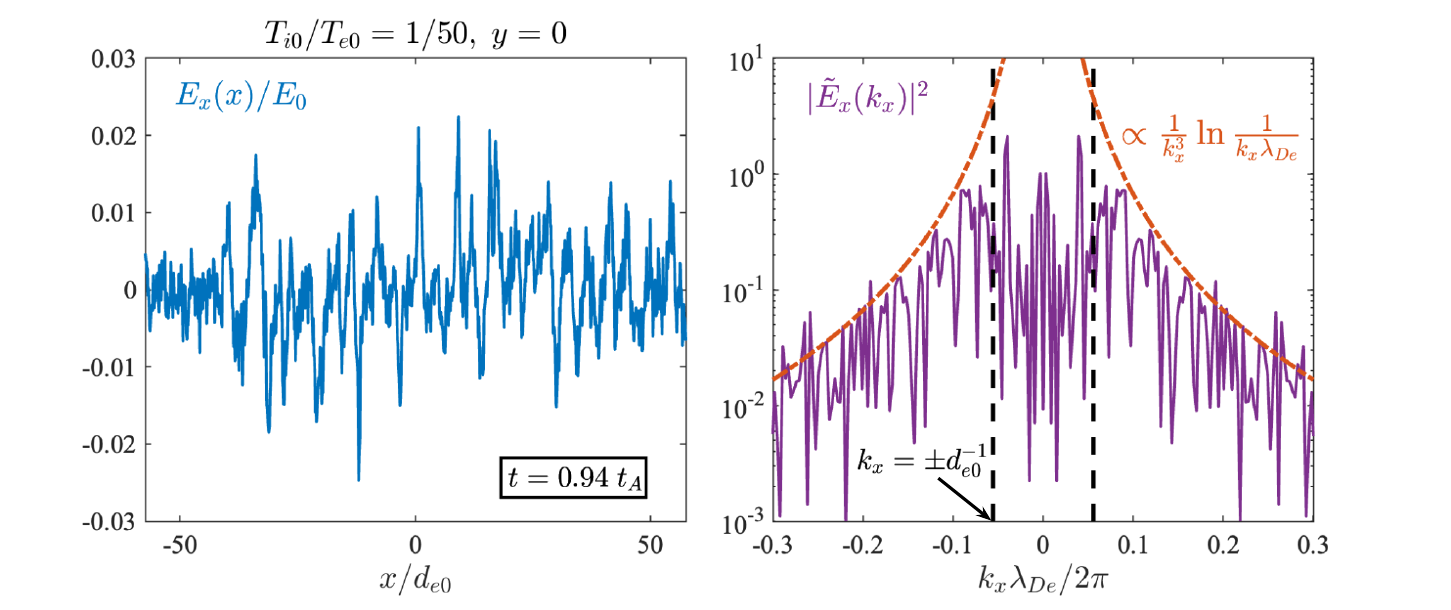}
\caption{\label{fig:wavenumspec} Outflow electric field (left) and the wavenumber spectrum (right) along the outflow symmetry line ($y = 0$) for $T_{i0}/T_{e0} = 1/50$ some time shortly after the peak reconnection rate. The wavenumber spectrum is compared with the KP spectrum (dot-dashed curve) and the approximate wavenumber of the fastest-growing IAI mode, roughly equal to the reciprocal of the electron skin-depth (see Fig.~\ref{fig:pk}), is indicated by the dashed lines. }
\end{figure*}

Beyond the alignment with the linear Vlasov-Poisson solution and the MHD compressional mode, the spectra in Fig.~\ref{fig:ss_y0} also show low‐frequency modes between the $\omega = 0$ branch (Eq.~\ref{subeq:MHDsol2}) and the solutions from Eq.~\ref{eq:solVlasov}. These modes emerge more clearly when performing one‐dimensional transforms of the outflow electric field near the x‐point, as seen in the right panel of Fig.~\ref{fig:wavenumspec}. 
Remarkably, the resulting one‐dimensional spectrum $|\tilde{E}_x (k_x)|^2$ along the outflow symmetry line agrees closely with the Kadomtsev-Petviashvili (KP) spectrum \citep{Kadomtsev1962,Petviashvili1963,Liu2024} suggesting that, in our simulations, saturation of IAI occurs through strong nonlinear ion effects (induced scattering) --- and, indeed, there is clear evidence for ion phase-space holes in Fig.~\ref{fig:ps}.

We conclude from the analysis of this section that the nonlinear wave activity that we observe in the current sheet is most likely induced by IAI. In the following sections we analyze the effects of this instability on the reconnection process.

\subsection{Effects of IAI on magnetic reconnection}\label{sec:effects_in_magnetic_reconnection}
This section examines the effects of IAI on collisionless magnetic reconnection, focusing first on the ion heating facilitated by IAI and subsequently on the anomalous contributions to the ion and electron momentum equations. 

\subsubsection{Ion heating}\label{sec:ionheating}
\begin{figure*}
\includegraphics[width=1\textwidth,scale=.5]{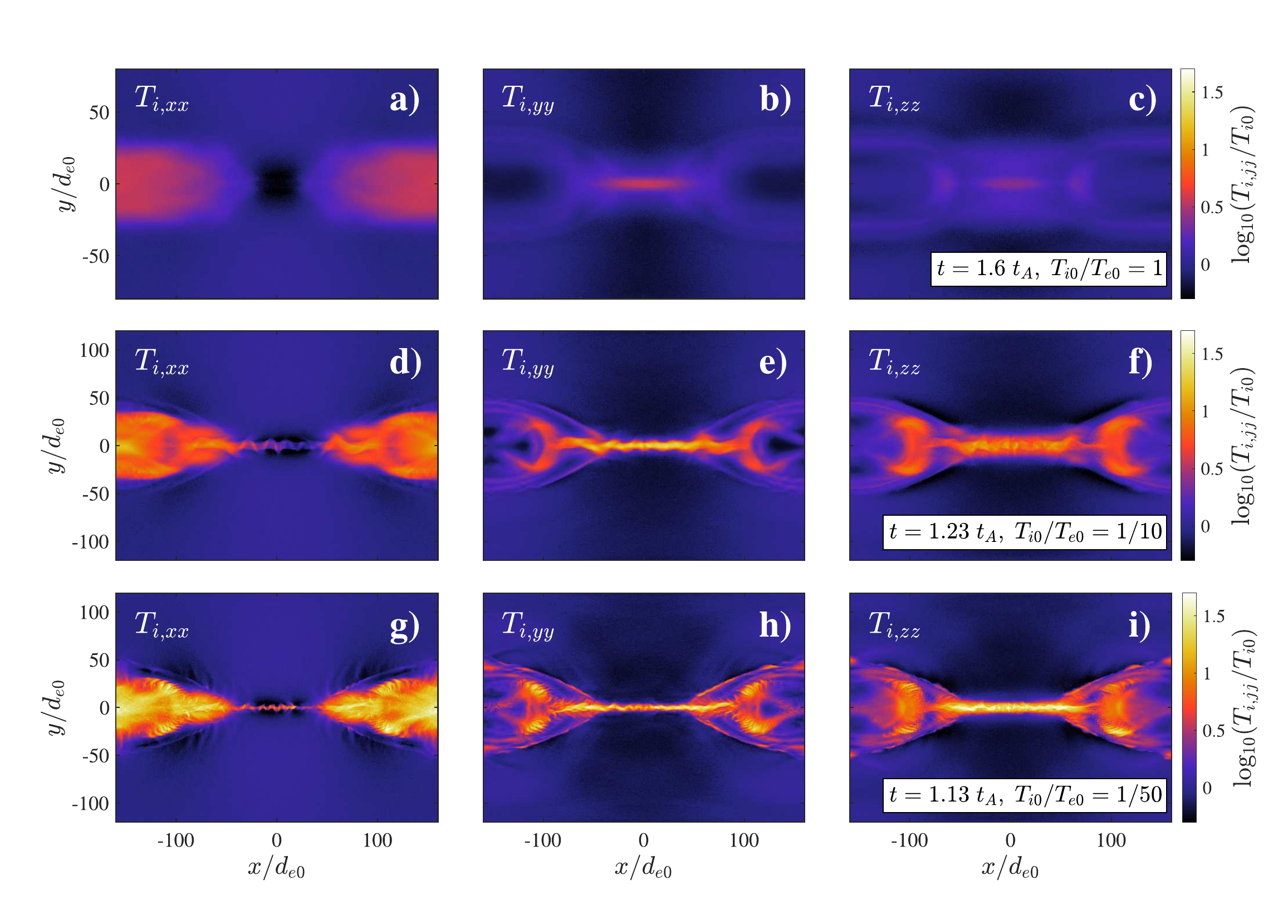}
\caption{\label{fig:Ti_tensor_log} Comparison of the logarithm (base 10) of the ion temperature tensor elements $T_{xx}$ (left column), $T_{yy}$ (middle column), and $T_{zz}$ (right column), normalized to the initial ion temperature $T_{i0}$, some time after the peak reconnection rate for $T_{i0}/T_{e0} = 1$ (top row), $1/10$ (middle row), and $1/50$ (bottom row). }
\end{figure*}

\begin{figure*}
\includegraphics[width=1\textwidth,scale=.5]{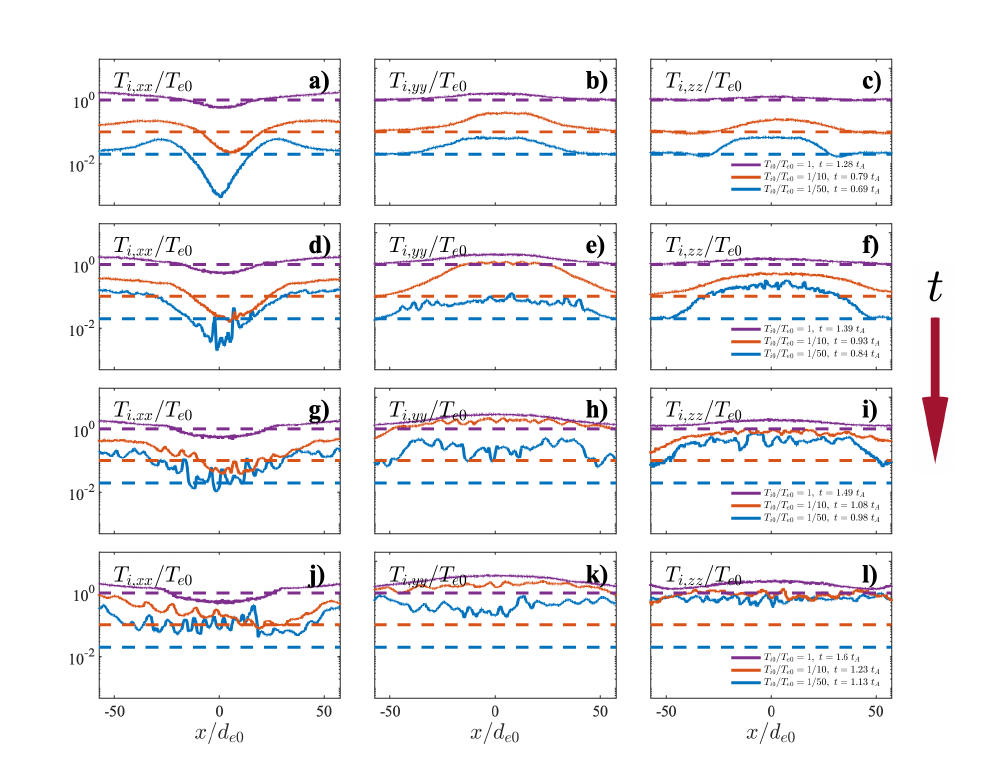}
\caption{\label{fig:Tijj_outflowsym} Comparison of the ion temperature tensor elements $T_{xx}$ (left column), $T_{yy}$ (middle column), and $T_{zz}$ (right column), normalized to the initial electron temperature $T_{e0}$, along the outflow symmetry line ($y = 0$), near the x-point, at times before (upper rows) and after (bottom rows) the peak reconnection rate for $T_{i0}/T_{e0} = 1$ (in purple), $1/10$ (in orange), and $1/50$ (in blue). The dashed lines in each subplot indicate the initial temperature ratio corresponding to its color. }
\end{figure*}

Our simulations reveal substantial ion heating in cold ion runs, particularly in the $xx$ component of the ion temperature tensor, as demonstrated in Figs.~\ref{fig:Ti_tensor_log} and~\ref{fig:Tijj_outflowsym}, which show the diagonal elements of the ion temperature tensor, normalized to $T_{i0}$, over the entire simulation domain, and their evolution along the outflow symmetry line, respectively. Due to reconnection dynamics, $T_{i,xx}$ undergoes an initial cooling to below $T_{i0}$ near the x-point (see Appendix~\ref{appB}). Subsequently, when the initial ratio is $T_{i0}/T_{e0} = 1/10$ and $1/50$ (and, thus, IAI is strongly triggered), $T_{i,xx}$ rises to $\approx T_e/10$ throughout the current sheet (while the electrons remain essentially unheated, so $T_e \approx T_{e0}$). In contrast, no comparable heating is observed in the equal-temperature run ($T_{i0}/T_{e0} = 1$), where $T_{i,xx}$ remains below $T_{i0}$ at the x-point. 

Although we have so far focused on the onset of in-plane IAI and the generation of IAWs in just the outflow direction, we similarly expect IAI to be triggered in the inflow direction, but more weakly so because the inflow electron-ion drift speeds $|U_{d,\text{inflow}}| = |u_{y,e} - u_{y,i}|$ are smaller than $|U_{d,\text{outflow}}|$ by about a factor of the reconnection rate $\mathcal{R}$. Furthermore, we already expect some heating in $T_{i,yy}$ during the onset of reconnection (see Appendix~\ref{appB}), which elevates the threshold drift speed needed to trigger IAI. Still, we find that there are indeed regions in the vicinity of the current sheet where $|U_{d,\text{inflow}}| > U_{d,\text{IAI}}$ (see Appendix~\ref{appC}) which likely explains the added increases to $T_{i,yy}$ after IAI is triggered near the x-point. 

Finally, Figs.~\ref{fig:Ti_tensor_log} and~\ref{fig:Tijj_outflowsym} likewise show that $T_{i,zz}$ reaches values comparable to $T_e$. We do not, however, attribute this heating to IAI but rather to the reconnection electric field $E_z$ (see Appendix~\ref{appB}). 
During the late stages of IAI, we might expect $T_{i,zz}$ (along with the other diagonal elements of the ion temperature tensor) to additionally pick up small-scale fluctuations, due to the highly non-Maxwellian nature of the velocity distributions at these times (see Fig.~\ref{fig:ps}), but IAI itself is not the primary mechanism for raising $T_{i,zz}$.

\subsubsection{Anomalous contributions}\label{sec:anomalous}
The presence of small-scale fluctuations in the current sheet allows for a (standard) decomposition of the fields into mean and fluctuating parts, and the computation of so-called anomalous terms in the momentum equations,
\begin{gather}
\mathbf{D}_{\alpha} \equiv - \frac{\langle \delta n_{\alpha} \delta \mathbf{E} \rangle}{\langle n_{\alpha} \rangle}, \label{eq:ac1_first} \\
\mathbf{T}_{\alpha} \equiv - \frac{\langle n_{\alpha} (\mathbf{u}_{\alpha} \times \mathbf{B}) \rangle}{\langle n_{\alpha} \rangle c} + \frac{\langle \mathbf{u}_{\alpha} \rangle \times \langle \mathbf{B} \rangle}{c}, \label{eq:ac2_first} \\
\mathbf{I}_{\alpha} \equiv \frac{m_{\alpha}}{q_{\alpha} \langle n_{\alpha} \rangle} \big[ \langle \nabla \cdot (\mathbf{u}_{\alpha} \mathbf{u}_{\alpha} n_{\alpha}) \rangle - \nabla \cdot (\langle \mathbf{u}_{\alpha} \rangle \langle \mathbf{u}_{\alpha} \rangle \langle n_{\alpha} \rangle) \big], \label{eq:ac3_first} \\
\mathbf{V}_{\alpha} \equiv \frac{m_{\alpha}}{q_{\alpha} \langle n_{\alpha} \rangle} \frac{\partial}{\partial t} \langle \delta n_{\alpha} \delta \mathbf{u}_{\alpha} \rangle, \label{eq:ac4_first}
\end{gather}
by performing two-dimensional spatial averaging in the plane of reconnection (see Appendix~\ref{appD} for the derivation of the anomalous terms and a detailed description of the averaging procedure). Here, $\mathbf{D}_{\alpha}$ is the anomalous drag (or resistivity), $\mathbf{T}_{\alpha}$ is the anomalous viscosity (momentum transport), $\mathbf{I}_{\alpha}$ is the anomalous Reynolds stress, and $\mathbf{V}_{\alpha}$ is associated with fluctuating flows \citep{Che2011,Graham2022}.

\begin{figure*}
\includegraphics[width=1\textwidth,scale=.5]{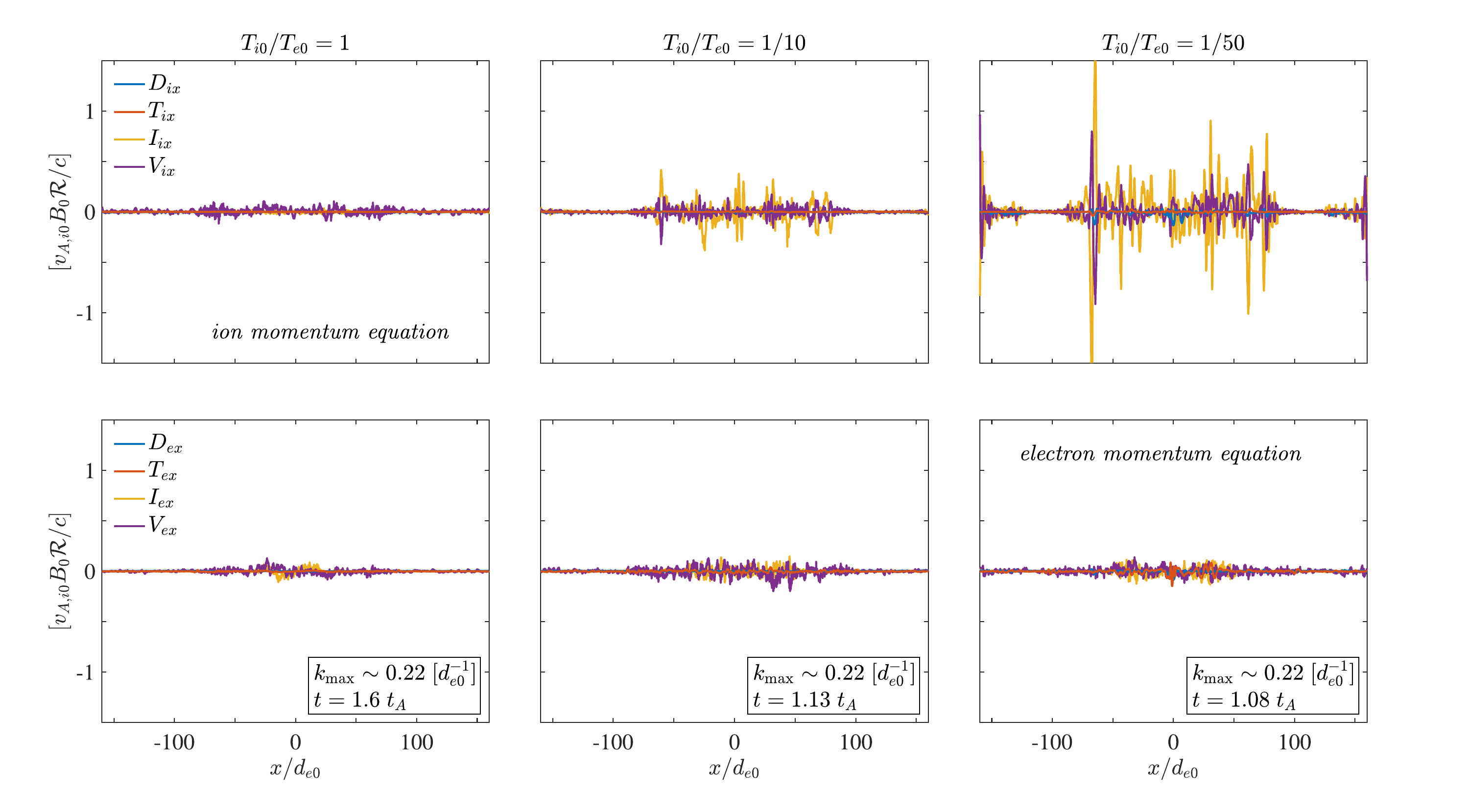}
\caption{\label{fig:iad_outflowsym} Comparison of the anomalous contributions (Eqs.~\ref{eq:ac1_first} to~\ref{eq:ac4_first}), normalized to $v_{A,i0} B_0 \mathcal{R}/c$ with $\mathcal{R} = 0.2$, obtained from simulations using the ion (top row) and electron (bottom row) momentum equations along the outflow symmetry line ($y = 0$) some time after the peak reconnection rate for $T_{i0}/T_{e0} = 1$ (left column), $1/10$ (middle column), and $1/50$ (right column). The maximum allowed $k$ indicated at the lower right corner of each subplot in the bottom row corresponds to the reciprocal of the length of the window size in which the Savitzky-Golay filter is implemented (see Appendix~\ref{appD}). }
\end{figure*}

Figure~\ref{fig:iad_outflowsym} shows the anomalous contributions obtained from simulations by spatially averaging the momentum equation along the outflow symmetry line some time after the peak reconnection rate. 
We normalize the anomalous terms to $v_{A,i0} B_0 \mathcal{R}/c$, where we have taken the (normalized) reconnection rate $\mathcal{R}$ to be approximately $0.2$ (guided by the data in Fig.~\ref{fig:ER_max}). 
As shown in these figures, we find that for systems with low initial ion-electron temperature ratios, anomalous Reynolds stress dominates the anomalous contributions to the ion momentum equation. Although the overall reconnection rate remains largely unaffected by IAI (see Fig.~\ref{fig:ER_max}), the localized dynamics in the diffusion region demonstrate significant fluctuations in both the Reynolds stress (as seen in $I_{ix}$ spikes in Fig.~\ref{fig:iad_outflowsym}) and the anomalous flow term $V_{ix}$. These spikes suggest the emergence of small-scale shear flows and turbulence driven by IAI-induced density and flow fluctuations, which can potentially explain the fluctuations in the reconnection rate shown in Fig.~\ref{fig:ER_max} after reconnection rate peaks and IAI is strongly triggered for $T_{i0}/T_{e0} = 1/10$ and $1/50$\footnote{The strong $I_{ix}$ and $V_{ix}$ spikes in Fig.~\ref{fig:iad_outflowsym} are shown to be comparable to $v_{A,i0} B_0 \mathcal{R}/c$ for cold-ion runs, which is roughly on the order of the $(\mathbf{u}_i \times \mathbf{B})_x/c = - u_{iz} B_y/c$ term in the ion momentum equation at the outflows and approximately equal to the magnitude of the reconnection electric field. These anomalous contributions may then have an appreciable effect on the bulk ion outflow speed $u_{ix}$, which presents itself as fluctuations in the reconnection rate (as seen in Fig.~\ref{fig:ER_max}). Indeed, these fluctuations occur on timescales compatible with the $E_x$ waveforms seen in Fig.~\ref{fig:ss_y0}. However, since these anomalous terms are negligible compared to $v_{A,i0} B_0 \mathcal{R}/c$ when averaged over the diffusion region, they cannot significantly affect the global behavior of the reconnection rate.}. In contrast, anomalous contributions to electron momentum equation are negligible compared to $v_{A,i0} B_0 \mathcal{R}/c$. Importantly, when averaged over the diffusion region, all anomalous terms—both to the ion and electron momentum equations—are nearly zero, suggesting that IAI does not significantly affect coherent $d_i$-scale structures through these anomalous terms.

Previous numerical studies have reported that IAI may produce significant anomalous resistivity \citep{Watt2002,Petkaki2003,Petkaki2006,Hellinger2004}. However, these studies often begin with super-critical configurations that can artificially amplify wave energy, potentially explaining the discrepancy between earlier results and our findings, which seem to indicate that although IAI causes substantial ion heating, it does not contribute to large-scale anomalous effects, at least in the context of magnetic reconnection. This diverges from the conclusions of \citet{Rudakov1966,Bychenkov1988} because our system never reaches a hypothetical steady-state regime in which ion‑acoustic turbulence drags enough momentum from the electrons to reduce the in‑plane current. In \citet{Rudakov1966,Bychenkov1988}, the electron-ion drift is assumed to stabilize near the ion‑sound speed once the instability saturates, which is only true if just the resonant electrons are taken into account. The bulk of the fast tail, which carries most of the current, remains non‑resonant and is freely accelerated due to reconnection dynamics (e.g., magnetic tension forces, electron pressure gradient). Consequently, the anomalous resistivity in our simulations is much lower than the theoretical estimate—a result that aligns with recent analyses of IAI \citep{Liu2024}. 

\section{Conclusions}\label{sec:conclusions}
This study presents comprehensive numerical evidence for the excitation of ion acoustic instability (IAI) within the diffusion region of a reconnecting magnetic configuration, underscoring the important role of this kinetic instability in influencing reconnection dynamics in environments where the upstream ion-electron temperature ratio is significantly below unity. 

In cases with low ion-electron temperature ratios, our results show that IAI induces significant ion heating, such that the ion temperature along the reconnection outflow rises to $\approx T_e/10$. On the other hand, we find that IAI has only a limited impact on the anomalous contributions to the momentum equations and the overall reconnection rate: in contrast to earlier studies reporting substantial anomalous resistivity (under super-critical conditions), our findings indicate that the anomalous effects of IAI on momentum transport are minimal. Whether this remains true with more realistic parameters—e.g. much larger separation between the macroscopic and the kinetic length-scales, as well as between the speed of light and the ion Alfv\'en speed—is an open question. 

We speculate that these results may have broader implications for ion heating in the solar wind, as it is understood that solar wind turbulence is pervaded by numerous reconnection sites \citep{Retin2007,Servidio2009,Servidio2011,Osman2014}, and the partitioning of energy at these sites could be partially governed by instabilities like IAI. 
As discussed in Section~\ref{sec:introduction}, spacecraft observations have consistently reported the existence of solar wind patches with low ion-electron temperature ratios and substantial but poorly understood ion heating across vast distances in the solar wind. One suggested explanation for such heating is that it may be due to IAI-mediated energy conversion in reconnection sites, spontaneously arising via turbulent dynamics. We argue in Appendix~\ref{appE} that the standard description of turbulence at sub-ion scales in the solar wind allows for a significant range of scales where indeed one might expect IAI to be triggered.

While our focus in this analysis is primarily on waveform generation near the x-point due to large electron-ion drifts, it is important to note that similar drift-driven wave activity is also observed along and near the separatrix (somewhat visible in Fig.~\ref{fig:comp}). The electron-ion drifts in these regions are similarly large, leading to the generation of waveforms and the further heating of ions. However, as reported in previous numerical studies, kinetic processes other than IAI likely contribute to the observed wave activity in those regions, such as two-stream or counter-streaming instabilities \citep{Fujimoto2014,Chen2015,Hesse2018} and the Kelvin-Helmholtz instability \citep{Divin2012,Fermo2012,Lapenta2014}. These additional instabilities near the separatrix play a complementary role in shaping the broader reconnection environment. Future studies could explore the interaction of these instabilities further to develop a more comprehensive understanding of the complex kinetic processes governing collisionless magnetic reconnection.

\section*{Acknowledgments}\label{sec:acknowledgments}
The authors extend their gratitude to Kyle G. Miller for his valuable insights on PIC simulations and to Muni Zhou for engaging discussions. The authors also acknowledge the OSIRIS Consortium, comprising UCLA and Instituto Superior T\'ecnico (IST) in Lisbon, Portugal, for providing access to the OSIRIS 4.0 framework. 

\section*{Funding}\label{sec:funding}
This research was partially funded by the National Science Foundation through the Graduate Research Fellowship Program (NSF GRFP) award and by DOE award DE-SC0022012. Additionally, this research utilized resources of the MIT-PSFC partition of the Engaging cluster at the MGHPCC facility, funded by DOE award No. DE-FG02-91-ER54109 and the National Energy Research Scientific Computing Center, a DOE Office of Science User Facility, supported by the U.S. Department of Energy under Contract No. DE-AC02-05CH11231 using NERSC award FES-ERCAP0026577.

\section*{Declaration of interests}\label{sec:declaration_of_interests}
The authors report no conflict of interest.

\section*{Data availability statement}\label{sec:data_availability_statement}
The data that support the findings of this study are available from the corresponding author upon reasonable request.

\appendix

\section{Temperature dependence of the IAI wavenumber associated with a marginally stable system}\label{appA}
In Fig.~\ref{fig:pk}, we illustrate that the wavenumber associated with a marginally stable system ($\gamma_{\text{max}} = 0$) increases sharply as $T_{i0}/T_{e0}$ decreases below about $1/10$. This behavior arises in the regime where $U_d \ll v_{\text{th},e}$ and $T_{i} \ll T_{e}$. In that limit, $c_s = \sqrt{(\gamma_e T_e + \gamma_i T_i)/m_i} \simeq \sqrt{T_e/m_i} \gg v_{\text{th},i}$ ($\gamma_e = 1$ for isothermal electrons). Thus at the ion-acoustic resonance ($\omega \simeq kc_s$), one obtains
\begin{eqnarray}
|\zeta_i| \equiv |\omega/(\sqrt{2}kv_{\text{th},i})| \gg 1, \label{eq:zetai}
\end{eqnarray}
and
\begin{eqnarray}
|\zeta_e| \equiv |(\omega - \mathbf{k} \cdot \mathbf{U}_d)/(\sqrt{2}kv_{\text{th},e})| \ll 1. \label{eq:zetae}
\end{eqnarray}
Substituting these conditions into Eq.~\ref{eq:solVlasov} then leads to \citep{Fitzpatrick2016}
\begin{eqnarray}
2 (k \lambda_{De})^2 \simeq \frac{T_e}{T_i} \frac{1}{\zeta_i^2} - 2 - 2i\sqrt{\pi} \bigg( \frac{T_e}{T_i} \zeta_i e^{-\zeta_i^2} + \zeta_e \bigg), \label{eq:Fitzpatrick}
\end{eqnarray}
and approximately \citep{Biskamp2000}
\begin{eqnarray}
[\Re (\omega)]^2 \simeq \frac{k^2 c_s^2}{1 + k^2 \lambda_{De}^2}, \label{eq:Biskamp}
\end{eqnarray}
assuming that $\gamma \equiv \Im (\omega) \ll \Re (\omega)$. 

It has been shown \citep{Fried1961} that the critical drift velocity for IAI is approximately $U_{d,\text{IAI}} \simeq 4\sqrt{2} v_{\text{th},i}$ when $T_i/T_e < 1/20$, and is of order $c_s$ for higher temperature ratios (e.g., $1/20 \lesssim T_i/T_e \lesssim 1$). Combining this with Eqs.~\ref{eq:Fitzpatrick} (upon ignoring the exponential term) and~\ref{eq:Biskamp} yields the final result
\begin{eqnarray}
k^2 (\gamma_{\text{max}} = 0) \lambda_{De}^2 \simeq
\begin{cases} 
(32 T_i/T_e)^{-1} - 1, & T_i/T_e \lesssim 1/20 \\
0, & 1/20 \lesssim T_i/T_e \lesssim 1
\end{cases}. \label{eq:me}
\end{eqnarray}
For $T_i/T_e = 1/50$, Eq.~\ref{eq:me} gives $k \approx 0.12~[2\pi/\lambda_{De}]$, which is in close agreement with the solution presented in Fig.~\ref{fig:pk} ($k_{\text{max}} (\gamma_{\text{max}} = 0) \approx 0.09~[2\pi/\lambda_{De}]$). Equation~\ref{eq:me} makes it evident that $k(\gamma_{\text{max}} = 0)$ increases with decreasing $T_i/T_e$, and also clarifies why $k_{\text{max}}$ at $\gamma_{\text{max}} = 0$ becomes (or remains) zero when $T_i/T_e$ increases beyond $1/10$ (see Fig.~\ref{fig:pk}). It is interesting to note that Eq.~\ref{eq:me} does not involve the ion-electron mass ratio, implying that when IAI just begins to become unstable, the same behavior of $k_{\text{max}}$ would emerge even if a more realistic ratio (e.g., $m_i/m_e = 1836$) were used.

\section{Diagonal ion temperature tensor elements at the x-point in the absence of IAI}\label{appB}
Taking the second moment of the Vlasov equation yields the evolution of the pressure tensor~\citep[e.g.][]{Ng2020b}
\begin{eqnarray}
\frac{\partial \mathcal{P}_{lm}}{\partial t} + \frac{\partial \mathcal{Q}_{lmn}}{\partial x_n} = nqu_{[l}E_{m]} + \frac{q}{m} \epsilon_{[lnp} \mathcal{P}_{nm]} \frac{B_p}{c}, \label{eq:secondmoment}
\end{eqnarray}
where the square brackets denote a sum over permutations of the indices (e.g., $u_{[l}E_{m]} = u_l E_m + u_m E_l$), $\mathcal{P}_{lm}$ and $\mathcal{Q}_{lmn}$ are the second and third moments of the distribution function,
\begin{subequations}
\begin{eqnarray}
\mathcal{P}_{lm} = m \int d^3 v~v_l v_m f = P_{lm} + mnu_l u_m,
\end{eqnarray}
\begin{eqnarray}
\mathcal{Q}_{lmn} = m \int d^3 v~v_l v_m v_n f = q_{lmn} + u_{[l}\mathcal{P}_{mn]} - 2mnu_l u_m u_n,
\end{eqnarray}
\end{subequations}
and $q_{lmn}$ is the heat flux tensor. 

The ion momentum equation is
\begin{eqnarray}
\mathbf{E}^{\prime} = \frac{m_i}{e n_i} \bigg[ \frac{\partial}{\partial t} (n_i \mathbf{u}_i) + \nabla \cdot (\mathbf{u}_i \mathbf{u}_i n_i) \bigg] + \frac{1}{e n_i} \nabla \cdot \mathbb{P}_i, \label{eq:Ohm}
\end{eqnarray}
where $\mathbf{E}^\prime = \textbf{E} + \textbf{u}_i \times \textbf{B}/c$ is the nonideal electric field, $\mathbb{P}_i$ is the ion pressure tensor, and the resistivity is neglected for collisionless reconnection. 

At the x-point, due to symmetry, it is approximately true that
\begin{subequations}
\begin{eqnarray}
\partial_y u_{ix} \big|_{X} = \partial_x u_{iy} \big|_{X} = u_{ix,X} = u_{iy,X} = 0, \label{subeq:profassumpt0}
\end{eqnarray}
\begin{eqnarray}
\nabla u_{iz} |_{X} = \mathbf{B}_X = \pmb{0}. \label{subeq:profassumpt1}
\end{eqnarray}
\end{subequations}
Assuming steady-state ($\partial_t = 0$) and uniform density near the x-point, the $xx$, $yy$, and $zz$ components of Eq.~\ref{eq:secondmoment}, the $z$ component of Eq.~\ref{eq:Ohm}, and Eqs.~\ref{subeq:profassumpt0} and~\ref{subeq:profassumpt1} give
\begin{subequations}
\begin{eqnarray}
\partial_x q_{i,xxx} \big|_X + \partial_y q_{i,xxy} \big|_X + n_{i,X} T_{i,xx,X} \big( 3 \partial_x u_{ix} \big|_X + \partial_y u_{iy} \big|_X \big) = 0, \label{subeq:secondmoment_xx}
\end{eqnarray}
\begin{eqnarray}
\partial_x q_{i,yyx} \big|_X + \partial_y q_{i,yyy} \big|_X + n_{i,X} T_{i,yy,X} \big( \partial_x u_{ix} \big|_X + 3 \partial_y u_{iy} \big|_X \big) = 0, \label{subeq:secondmoment_yy}
\end{eqnarray}
\begin{align}
\partial_x q_{i,zzx} \big|_X + \partial_y q_{i,zzy} \big|_X & + n_{i,X} \big[ ( T_{i,zz,X} + m_i u_{iz,X}^2 ) \big( \partial_x u_{ix} \big|_X + \partial_y u_{iy} \big|_X \big) \big] \nonumber \\
& + 2 u_{iz,X} \big( \partial_x T_{i,xz} \big|_X + \partial_y T_{i,yz} \big|_X \big) = 2e n_{i,X} E_{z,X} u_{iz,X}, \label{subeq:secondmoment_zz}
\end{align}
\begin{eqnarray}
e E_{z,X} = \partial_x T_{i,xz} \big|_X + \partial_y T_{i,yz} \big|_X + m_i u_{iz,X} \big( \partial_x u_{ix} \big|_X + \partial_y u_{iy} \big|_X \big), \label{subeq:Ohm_z}
\end{eqnarray}
\end{subequations}
which may be rearranged to read
\begin{subequations}
\begin{eqnarray}
T_{i,xx,X} = - \frac{1}{n_{i,X}} \frac{\partial_x q_{i,xxx} \big|_X + \partial_y q_{i,xxy} \big|_X}{3 \partial_x u_{ix} \big|_X + \partial_y u_{iy} \big|_X}, \label{subeq:Tixx_xline}
\end{eqnarray}
\begin{eqnarray}
T_{i,yy,X} = - \frac{1}{n_{i,X}} \frac{\partial_x q_{i,yyx} \big|_X + \partial_y q_{i,yyy} \big|_X}{\partial_x u_{ix} \big|_X + 3 \partial_y u_{iy} \big|_X}, \label{subeq:Tiyy_xline}
\end{eqnarray}
\begin{eqnarray}
T_{i,zz,X} = m_i u_{iz,X}^2 - \frac{1}{n_{i,X}} \frac{\partial_x q_{i,zzx} \big|_X + \partial_y q_{i,zzy} \big|_X}{\partial_x u_{ix} \big|_X + \partial_y u_{iy} \big|_X}. \label{subeq:Tizz_xline}
\end{eqnarray}
\end{subequations}

We note that at the x-point, ions experience the reconnection electric field
\begin{eqnarray}
E^\prime_{z,X} = E_{z,X} \approx - \frac{v_{A,i} B_0 \mathcal{R}}{c}, \label{subeq:Ezprime}
\end{eqnarray}
for a duration $\Delta t \sim \delta/v_{\text{in}}$, where $\delta \sim d_i \sim d_{i0}$ is the thickness of the ion diffusion region, $v_{\text{in}} \sim \mathcal{R} v_{A,i}$ is the inflow speed, and $\mathcal{R}$ is the reconnection rate. We may therefore estimate that at the x-point, the out-of-plane ion velocity increases by
\begin{eqnarray}
u_{iz,X} - u_{iz0} \sim \frac{e}{m_i} E^\prime_{z,X} \Delta t \sim - v_{A,i}, \label{eq:Deltaviz}
\end{eqnarray}
where in our simulations, $u_{iz0}$ is specified by the choices $\omega_{pe}/\Omega_{ce} = 2$ and $\lambda_{\text{B}}/d_{i0} = 1$, and the force-balance and Vlasov equilibrium conditions (see Section~\ref{sec:simulation_setup}). It is written as
\begin{eqnarray}
u_{iz0} = - \frac{c}{2} \sqrt{\frac{m_e}{m_i}} \bigg( \frac{T_{e0}}{T_{i0}} + 1 \bigg)^{-1}. \label{eq:uiz0}
\end{eqnarray}
Utilizing Eqs.~\ref{subeq:Tixx_xline} to~\ref{subeq:Tizz_xline},~\ref{eq:Deltaviz}, and~\ref{eq:uiz0} and considering a zero heat flux leads to
\begin{subequations}
\begin{eqnarray}
T_{i,xx,X} = T_{i,yy,X} = 0, \label{subeq:Tixx_xline_zeroheatflux}
\end{eqnarray}
\begin{eqnarray}
T_{i,zz,X} \sim \frac{2 (T_{e0} + 2 T_{i0})^2}{T_{e0} + T_{i0}}, \label{subeq:Tizz_xline_zeroheatflux}
\end{eqnarray}
\end{subequations}
where, in writing the right-hand relation of Eq.~\ref{subeq:Tizz_xline_zeroheatflux}, we have assumed a plasma $\beta$ of order unity ($8\pi n (T_{i0} + T_{e0}) \sim B_0^2$). It is evident in Eq.~\ref{subeq:Tizz_xline_zeroheatflux} that for $T_{i0}/T_{e0} \ll 1$, $T_{i,zz,\text{max}} \sim 2 T_{e0} \sim T_{e0}$, which explains that the out-of-plane heating we observe in Figs.~\ref{fig:Ti_tensor_log} and~\ref{fig:Tijj_outflowsym} is due primarily to the reconnection electric field $E_z$ and implies that we expect $T_{i,zz}$ to reach $\sim T_e$ even in the absence of IAI.

The fact that Eqs.~\ref{eq:secondmoment},~\ref{subeq:profassumpt0}, and~\ref{subeq:profassumpt1} formally lead to the solution Eq.~\ref{subeq:Tixx_xline_zeroheatflux} must be recognized as a failure of the fluid model (that is, of the zero heat-flux closure assumed in the derivation). Still, Eq.~\ref{subeq:Tixx_xline_zeroheatflux} suggests that $T_{i,xx}$ must undergo a substantial drop from $T_{i0}$ at the x‐point—consistent with our simulations (see Figs.~\ref{fig:Ti_tensor_log} and~\ref{fig:Tijj_outflowsym}).

On the other hand, the zero heat flux assumption also implies a drop in $T_{i,yy}$ below $T_{i0}$ at the x-point, which sharply contradicts our simulation results showing appreciable $T_{i,yy,X}/T_{i0}$, even in runs where IAI does not occur. One reason for this discrepancy could be that such a closure might be more appropriate for $T_{i,xx}$ than for $T_{i,yy}$ because unlike $T_{i,xx}$, ions entering the current sheet gain significant $T_{i,yy}$ in regions of large inflow velocity $u_{iy}$ and could maintain that $T_{i,yy}$ upon arrival at the x‐point. We can estimate this local $T_{i,yy}$ increase by examining the approximate $yy$ temperature at the vertical edges of the current sheet (i.e., in the upstream direction), which might be on the same order of magnitude as $T_{i,yy,X}$. Along the inflow symmetry line, $x = 0$, symmetry arguments similar to Eqs.~\ref{subeq:profassumpt0} and~\ref{subeq:profassumpt1} can be made
\begin{align}
\partial_y u_{ix} \big|_{x = 0} & = \partial_x u_{iy} \big|_{x = 0} = \partial_x u_{iz} \big|_{x = 0} \nonumber \\
& = u_{ix} (x = 0) = B_{y} (x = 0) = B_{z} (x = 0) = 0, \label{eq:profassumpt2}
\end{align}
which applied to the $yy$ and $zz$ components of Eqs.~\ref{eq:secondmoment} and~\ref{eq:Ohm} yields the relations
\begin{subequations}
\begin{eqnarray}
u_{iy} [ -2eE_y + 2 \partial_x T_{i,xy} + 3 \partial_y T_{i,yy} + m_i u_{iy} (\partial_x u_{ix} + 3 \partial_y u_{iy}) ] \nonumber \\
= 2 \Omega_{ix} (T_{i,yz} + m_i u_{iy} u_{iz}) - T_{i,yy} (\partial_x u_{ix} + 3 \partial_y u_{iy}), \label{subeq:secondmoment_yy2}
\end{eqnarray}
\begin{eqnarray}
u_{iz} [-2eE_z + 2 \partial_x T_{i,xz} + 2 \partial_y T_{i,yz} + 2 m_i u_{iy} \partial_y u_{iz} + m_i u_{iz} (\partial_x u_{ix} + \partial_y u_{iy})] \nonumber \\
= - 2 \Omega_{ix} (T_{i,yz} + m_i u_{iy} u_{iz}) - u_{iy} \partial_y T_{i,zz} \nonumber \\
- 2 T_{i,yz} \partial_y u_{iz} - T_{i,zz} (\partial_x u_{ix} + \partial_y u_{iy}), \label{subeq:secondmoment_zz2}
\end{eqnarray}
\begin{eqnarray}
\partial_x T_{i,xy} + \partial_y T_{i,yy} + m_i u_{iy} (\partial_x u_{ix} + 2 \partial_y u_{iy}) = eE_y + m_i \Omega_{ix} u_{iz}, \label{subeq:Ohm_y2}
\end{eqnarray}
\begin{eqnarray}
\partial_x T_{i,xz} + \partial_y T_{i,yz} + m_i u_{iy} \partial_y u_{iz} + m_i u_{iz} (\partial_x u_{ix} + \partial_y u_{iy}) \nonumber \\
= e E_z - m_i \Omega_{ix} u_{iy}, \label{subeq:Ohm_z2}
\end{eqnarray}
\end{subequations}
where $\Omega_{ix} \equiv eB_x/(m_i c)$ and we have made the simplifying assumption that heat flux away from the x-point (but not \textit{at} the x-point) is negligible, i.e., $\partial_n q_{lmn} = 0$.

Since we are considering steady-state and uniform density, the continuity equation gives the incompressibility condition
\begin{eqnarray}
\nabla \cdot \mathbf{u}_i = 0, \label{eq:incompressibility}
\end{eqnarray}
which combined with Eqs.~\ref{subeq:secondmoment_yy2} to~\ref{subeq:Ohm_z2} gives
\begin{eqnarray}
( - 2 u_{iy}^{-1} \partial_y u_{iy} - \partial_y ) T_{i,yy} = (1 + \Omega_{ix}^{-1} \partial_y u_{iz})^{-1} \partial_y T_{i,zz}, \label{eq:Tiyy_delta1}
\end{eqnarray}
along $x = 0$ and away from the x-point. We assume that
\begin{eqnarray}
(u_{iz,X} - u_{iz,\delta})^{-1} \partial_y u_{iz} \big|_{\delta} \approx - u_{iy,\delta}^{-1} \partial_y u_{iy} \big|_{\delta} \approx d_{i0}^{-1} \gtrsim - T_{i,yy,\delta}^{-1} \partial_y T_{i,yy} \big|_{\delta} > 0, \label{eq:profassumpt5}
\end{eqnarray}
where quantities with $\delta$ in the subscript indicates values at $x = 0$ and the $y$-ends of the current sheet, and, in writing the second to last inequality, we have noted that $T_{i,yy}$ might not undergo as drastic of a change from the inflow edges of the current sheet to the x-point compared to, e.g., $u_{iy}$ or $u_{iz}$ since $T_{i,yy,X}$ is not necessarily zero with the inclusion of non-zero heat flux. As mentioned above, we may approximately say
\begin{eqnarray}
T_{i,yy,X} \sim T_{i,yy,\delta}. \label{eq:Tiyy_assumption}
\end{eqnarray}
From Eq.~\ref{subeq:Tizz_xline}, we have
\begin{eqnarray}
\partial_y T_{i,zz} \big|_{\delta} \sim 2 m_i u_{iz,\delta} \partial_y u_{iz} \big|_{\delta}, \label{subeq:Tizz_xline2}
\end{eqnarray}
which combined with Eqs.~\ref{eq:Tiyy_delta1} and~\ref{eq:profassumpt5}, and the assumptions that $B_{x,\delta} \approx B_0$ and $u_{iz,\delta} \approx u_{iz0}$, gives
\begin{eqnarray}
4/3 \lesssim T_{i,yy,\delta}/T_{e0} \lesssim 2, \label{eq:Tiyy_delta2}
\end{eqnarray}
implying that we also expect substantial heating in $T_{i,yy}$ up to $\sim T_e$ in the vicinity of the x-point and in the absence of IAI. IAI triggered in the $y$-direction near the x-point might be responsible for additional increases in $T_{i,yy}$ after reconnection onset (see Appendix~\ref{appC}).

\section{$y$-component of in-plane IAI}\label{appC}
\begin{figure*}
\includegraphics[width=1.0\textwidth,scale=.5]{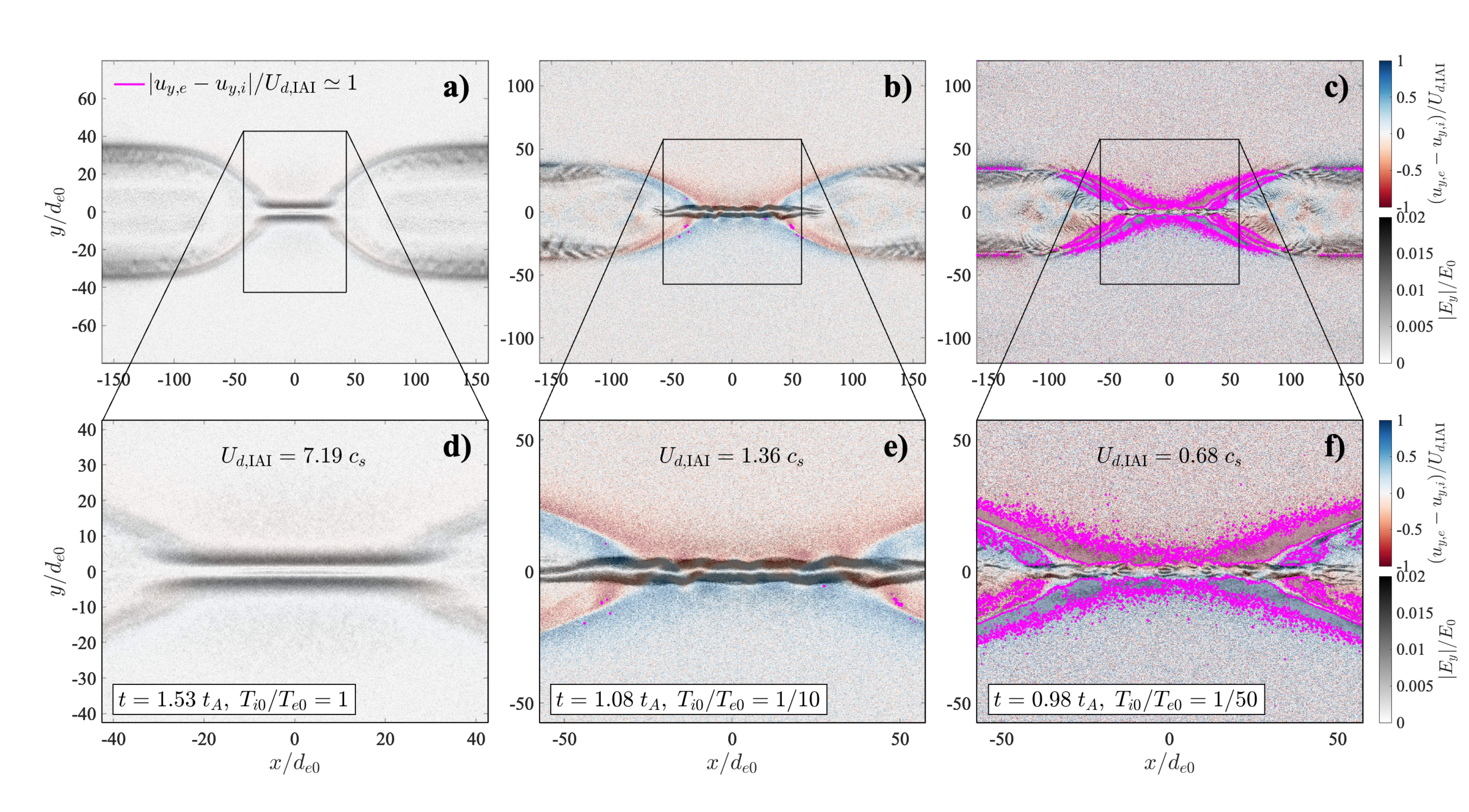}
\caption{\label{fig:compy} Colormap displaying the $y$-component of the electron-ion drift velocity normalized to the IAI threshold drifts, calculated with local ion-acoustic speeds $c_s = \sqrt{(T_{e,yy} + 3 T_{i,yy})/m_i}$ (red-blue), overlaid with contours indicating the regions where $|u_{y,e} - u_{y,i}|/U_{d,\text{IAI}} \simeq 1$ (magenta) and the magnitude of the inflow electric field ($E_y$, normalized to the reference value $E_0 = m_e c \omega_{pe}/e$ and in gray-scale), plotted at the same times as in Fig.~\ref{fig:comp}. This is shown for $T_{i0}/T_{e0} = 1$ (left-most column), $T_{i0}/T_{e0} = 1/10$ (middle column), and $T_{i0}/T_{e0} = 1/50$ (right-most column). 
The bottom row figures zoom in on the diffusion region, as identified by the boxes in the top row. }
\end{figure*}

Figure~\ref{fig:compy} illustrates the inflow electron-ion drift velocity and regions where this drift exceeds the IAI threshold drift, calculated with local ion-acoustic speeds $c_s = \sqrt{(T_{e,yy} + 3 T_{i,yy})/m_i}$, now with $T_{\alpha,yy}$ instead of $T_{\alpha,xx}$. Also shown is the magnitude of the inflow electric field ($E_y$) whose perturbations could be associated with inflow IAWs (although the overall structure of the electric field would mostly be influenced by perturbations due to the outflow IAI).

We mentioned in Section~\ref{sec:ionheating} that we expect the inflow IAI to be less strongly triggered than its outflow counterpart because inflow electron-ion drift speeds are smaller than outflow drift speeds (roughly by a factor of the reconnection rate $\mathcal{R}$) and that the early heating in $T_{i,yy}$ (see Appendix~\ref{appB}) raises the threshold to trigger IAI. Indeed, Fig.~\ref{fig:compy} shows that $|U_{d,\text{inflow}}|$ is only on the order of the IAI threshold drift in localized regions near the current sheet, whereas there are substantial regions along the outflow where $|U_{d,\text{outflow}}|$ significantly exceeds the threshold. Nevertheless, the existence of regions where $|U_{d,\text{inflow}}| > U_{d,\text{IAI}}$ implies that IAI plays a part in heating $T_{i,yy}$ for cases with cold ions.

\section{Derivation of anomalous terms and numerical averaging procedure}\label{appD}
We start from the momentum equation for species $\alpha$,
\begin{eqnarray}
m_{\alpha} \frac{\partial}{\partial t} (n_{\alpha} \mathbf{u}_{\alpha}) + m_{\alpha} \nabla \cdot (\mathbf{u}_{\alpha} \mathbf{u}_{\alpha} n_{\alpha}) = q_{\alpha} n_{\alpha} \bigg( \mathbf{E} + \frac{\mathbf{u}_{\alpha} \times \mathbf{B}}{c} \bigg) - \nabla \cdot \mathbb{P}_{\alpha},
\end{eqnarray}
where $\mathbb{P}_{\alpha}$ is the pressure tensor, and we have neglected collisions. 
Let us formally decompose each field into quasi-stationary (mean, averaged) and fluctuating components, such that $\mathcal{Q} = \langle \mathcal{Q} \rangle + \delta \mathcal{Q}$. 
We then find
\begin{eqnarray}
\frac{m_{\alpha}}{q_{\alpha} \langle n_{\alpha} \rangle} \bigg[ \frac{\partial}{\partial t} (\langle n_{\alpha} \rangle \langle \mathbf{u}_{\alpha} \rangle) + \nabla \cdot (\langle \mathbf{u}_{\alpha} \rangle \langle \mathbf{u}_{\alpha} \rangle \langle n_{\alpha} \rangle) \bigg] + \mathbf{V}_{\alpha} + \mathbf{I}_{\alpha} \nonumber \\
= \langle \mathbf{E} \rangle - \mathbf{D}_{\alpha} + \frac{\langle \mathbf{u}_{\alpha} \rangle \times \langle \mathbf{B} \rangle}{c} - \mathbf{T}_{\alpha} - \frac{\langle \nabla \cdot \mathbb{P}_{\alpha} \rangle}{q_{\alpha} \langle n_{\alpha} \rangle},
\end{eqnarray}
where $\mathbf{D}_{\alpha}$, $\mathbf{T}_{\alpha}$, $\mathbf{I}_{\alpha}$, and $\mathbf{V}_{\alpha}$ are defined in Eqs.~\ref{eq:ac1_first} to~\ref{eq:ac4_first}.

Expressions~\ref{eq:ac1_first} to~\ref{eq:ac4_first} are plotted in Fig.~\ref{fig:iad_outflowsym} for all $T_{i0}/T_{e0}$ considered at times after the reconnection rate peaks. For obtaining the numerical values in Fig.~\ref{fig:iad_outflowsym}, $\langle \mathcal{Q} \rangle$ was set as a two-dimensional spatial average in the plane of reconnection, performed using a Savitzky-Golay filter, which improves data precision while maintaining overall signal trends \citep{Savitzky1964}. The selection of $k_{\text{max}}$ (reciprocal of the length of the filtering window size) was carefully chosen to low-pass filter wavenumbers associated with the fastest-growing IAI modes and to reduce numerical noise. The upper bound of $k_{\text{max}}$ was established based on the noise in the $T_{i0}/T_{e0} = 1$ simulation. Despite this filtering, some residual noise remains visible (see the $T_{i0}/T_{e0} = 1$ plots in Fig.~\ref{fig:iad_outflowsym}), so it was necessary to select the smallest possible value for $k_{\text{max}}$ that would appropriately reduce this noise. Guided by the data in Fig.~\ref{fig:pk} (and Fig.~\ref{fig:wavenumspec}), which indicate that the wavenumbers of the fastest-growing IAI modes across all temperature ratios are approximately $\sim 0.05~[2\pi/\lambda_{De}] \sim 1~[d_{e0}^{-1}]$, we selected a $k_{\text{max}}$ smaller than this approximate wavenumber to ensure that all contributions from IAWs were filtered out.

Furthermore, our analysis shows that the initial ion-electron temperature ratio does not significantly affect the width of the current sheet. Across all simulations, the current sheet width (full width half maximum) at the time of peak reconnection rate remained consistent at approximately $\sim 5 d_{e0}$ or $0.5 d_{i0}$, allowing for the application of the same filtering technique for all temperature ratios with window sizes corresponding to $k_{\text{max}}^{-1}$.

\section{Possible connection to solar wind heating}\label{appE}
To connect our results with conditions in the solar wind, we adopt a standard model of kinetic‑Alfv\'en‑wave (KAW) turbulence \citep{Boldyrev2012,Zhou2023,Liu2025} to estimate how large the in-plane drift velocity due to turbulent reconnection events at sub-ion scales might become. 
We consider the range of scales $d_e \ll k_{\perp}^{-1} \ll \rho_i$, with $\rho_i = v_{\text{th},i}/\Omega_{ci}$ the ion Larmor radius, and invoke the usual scale-by-scale equipartition between density and magnetic perturbations. 
At transverse scale $\lambda \sim k_{\perp}^{-1}$, the electrostatic potential obeys $\varphi_{\lambda} \sim \sqrt{T_{i0}/T_{e0}} (\rho_i v_{A,i}/c) \delta B_{\perp \lambda}$ \citep{Liu2025}, where $\delta B_{\perp \lambda}$ is the field-perpendicular (fluctuating) magnetic field. Including Boldyrev-type intermittency \citep{Boldyrev2012}, the energy cascade rate (or eddy-turnover rate) for perpendicular fluctuations subject to nonlinear interactions is $\gamma_{nl} \sim 8\pi \varepsilon/(\delta B_{\perp \lambda}^2 p_{\lambda})$, where $\varepsilon$ denotes the (constant) energy flux, $p_{\lambda} = (\lambda/\lambda_0)^{3 - D} = \lambda/\lambda_0$ \citep{Frisch1995} is the volume-filling fraction of the two-dimensional structures lying in planes perpendicular to the background magnetic field, and the characteristic size $\lambda_0$ represents the largest scale of these highly intermittent, quasi-2D, field-perpendicular structures.

The in-plane IAI due to reconnection between two turbulent eddies becomes unstable when
\begin{eqnarray}
\frac{U_{d,\text{in-plane},\lambda}}{c_s} \sim \frac{v_{A,e,\perp \lambda}}{c_s} \sim \frac{\delta B_{\perp \lambda}}{\sqrt{4\pi n_0 m_e}} \sqrt{\frac{m_i}{T_{e0}}} \gtrsim 1. \label{eq:inplaneSWcriteria}
\end{eqnarray}
Combining $\gamma_{nl} \sim (c/B_0) \varphi_{\lambda}/\lambda^2$ with the above expressions, we obtain the range of scales where this can occur
\begin{eqnarray}
\frac{\lambda}{d_i} \gtrsim \frac{1}{\sqrt{8}} \frac{m_e^{3/2}}{m_i^2} \frac{n_0 \beta_e^{1/2}}{\varepsilon \lambda_0} T_{e0}^{1/2} T_{i0}, \label{eq:turbrange}
\end{eqnarray}
where $\beta_e \equiv 8\pi n_0 T_{e0}/B_0^2$ is the electron plasma beta and $B_0$ denotes the constant background (guide) magnetic field strength. 
The condition for the in-plane IAI to be triggered at \textit{any} scale ($d_e < \lambda < d_i$) then becomes
\begin{eqnarray}
T_{e0}^{1/2} T_{i0} \lesssim \frac{m_i^{3/2}}{m_e} \frac{\sqrt{8} \varepsilon \lambda_0}{n_0 \beta_e^{1/2}}. \label{eq:fluxcondition}
\end{eqnarray}
Observations of the solar wind by the Ulysses spacecraft \citep{Smith1995,Balogh1995}, suggest that the flux rate is $\varepsilon/(n_0 m_i) \sim 10^2$ J$~$kg$^{-1}$s$^{-1}$ \citep{SorrisoValvo2007}, which combined with established estimates of the solar wind electron plasma beta of $\beta_e \approx 1$ \citep{WilsonIII2018,Halekas2020}, and approximating $\lambda_0$ as the outer (energy-injection) perpendicular scale of solar-wind turbulence of $\lambda_C^{\perp} \sim 10^6$ km \citep{Weygand2009,Weygand2011} results in the condition Eq.~\ref{eq:fluxcondition} becoming $(T_{i0}^2 T_{e0})^{1/3} \lesssim 10^2$ eV, well above typical solar‑wind temperatures \citep{Richardson2003,Hellinger2013,tverk2015,Chen2016,Verscharen2019,Pa2021}.

Essential to the validity of the calculation above is to check if the reconnection rate is higher than the eddy turn over rate, or there would be no time for two eddies to reconnect before being sheared apart by the turbulence. 
Assuming that, at these scales, reconnection proceeds as described by the electron-only reconnection model of~\cite{Liu2025}, the reconnection time is $\tau_{\text{rec},\lambda} \sim \sqrt{2}\, (T_{e0}/T_{i0} + 1)^{-1/2} (\lambda/\rho_i) (\lambda/v_{A,i,\perp \lambda})$ and, therefore, 
\begin{eqnarray}
\gamma_{nl} \tau_{\text{rec},\lambda} \sim \frac{\sqrt{2} T_{i0}/T_{e0}}{\sqrt{1 + T_{i0}/T_{e0}}}, \label{eq:taurec}
\end{eqnarray}
which is small when the ion-electron temperature ratio $T_{i0}/T_{e0}$ is low, precisely the regime already required for the instability. 

An analogous estimate for the out‑of‑plane drift,
\begin{eqnarray}
\frac{U_{d,\text{out-of-plane},\lambda}}{c_s} \sim \frac{\delta J_{\lambda}/en_0}{c_s} \sim \frac{c}{4\pi} \frac{\delta B_{\perp \lambda}}{en_0 d_e} \sqrt{\frac{m_i}{T_{e0}}} \gtrsim 1,
\end{eqnarray}
recovers exactly the same scale constraint, Eq.~\ref{eq:turbrange}, and global threshold, Eq.~\ref{eq:fluxcondition}. Hence, under solar-wind parameters, both in-plane and out-of-plane IAIs should be readily triggered once thin current sheets reconnect.

It is worth noting that the above derivation assumes $|\delta B_{\perp}/B_0| \ll 1$, whereas our study focuses on a configuration without a guide field. Although the impact of a finite guide field on in-plane (perpendicular) and out-of-plane (parallel) IAI remains unclear, preliminary simulations including a guide field on the order of the in-plane, far-asymptotic field continue to show similar wave activity in the diffusion region. Future work could investigate in detail how a guide field influences in-plane IAI in the context of reconnection, and would be necessary to solidify the validity of the estimates made in this appendix.

\bibliographystyle{jpp}

\bibliography{main}

\end{document}